%% file: main.tex
\documentclass[fleqn,10pt]{wlscirep}
\usepackage[utf8]{inputenc}
\usepackage[T1]{fontenc}
\usepackage{algorithm}
 \usepackage{graphicx}
 \usepackage{subcaption}
 \usepackage{comment}
 \usepackage{pdfpages}
\usepackage{algpseudocode} 
\usepackage{siunitx}
\usepackage{amsmath}
\usepackage[section]{placeins}
\usepackage{bm}
\newcommand{\ovrlp}[2]{{\ensuremath{\langle #1|#2\rangle}}}
\newcommand{\ket}[1]{{\ensuremath{|#1\rangle}\xspace}}
\newcommand{\bra}[1]{{\ensuremath{\langle #1|}\xspace}}
\newcommand{\elemm}[3]{\bra{#1} #2 \ket{#3}}
\usepackage{braket}
\usepackage{caption}
\usepackage{lipsum}
\usepackage{tikz}
\usetikzlibrary{quantikz}
\usepackage{amsfonts}
\usepackage{amssymb}

\usepackage{soul}

\title{Overlap-ADAPT-VQE: Practical Quantum Chemistry on Quantum Computers via Overlap-Guided Compact Ansätze}

\author[1,2,+]{César Feniou}
\author[3,+]{Muhammad Hassan}
\author[1]{Diata Traoré}
\author[1]{Emmanuel Giner}
\author[3,4, *]{Yvon Maday}
\author[1,2,*]{Jean-Philip Piquemal}

\affil[1]{Sorbonne Universit\'e, Laboratoire de Chimie Théorique (UMR-7616-CNRS), F-75005 Paris, France}
\affil[2]{Qubit Pharmaceuticals, Paris, France}
\affil[3]{Sorbonne Université, CNRS, Université Paris Cité, Laboratoire Jacques-Louis Lions (LJLL), F-75005 Paris, France}
\affil[4]{Institut Universitaire de France, 75005, Paris, France}

\affil[*]{yvon.maday@sorbonne-universite.fr, ~jean-philip.piquemal@sorbonne-universite.fr}

\affil[+]{These authors contributed equally to this work}

\input{abstract.tex}
\begin{document}

\flushbottom
\maketitle
%
%
\thispagestyle{empty}

\section*{Introduction}
\input{introductionn.tex}

\section*{Technical Background and Methods}
\input{technical_background.tex}

\section*{Results}
\input{results2.tex}


 

\section*{Discussion}
\input{Discussion.tex}


\section*{Acknowledgements}

This work has been funded by the European Research Council (ERC) under the European Union’s Horizon 2020 research and innovation program (grant No 810367), project EMC2 (J.-P. P. and Y.M.)

\section*{Appendix}

\input{Appendix.tex}

\bibliography{achemso-demo.bib}









\end{document}

%% file: abstract.tex
\begin{abstract}
ADAPT-VQE is a robust algorithm for hybrid quantum-classical simulations of quantum chemical systems on near-term quantum computers. While its iterative process systematically reaches the ground state energy, practical implementations of ADAPT-VQE are sensitive to local energy minima, leading to over-parameterized ansätze. We introduce the Overlap-ADAPT-VQE to grow wave-functions by maximizing their overlap with any intermediate target wave-function that already captures some electronic correlation. By avoiding building the ansatz in the energy landscape strewn with local minima, the Overlap-ADAPT-VQE produces ultra-compact ansätze suitable for high-accuracy initialization of a new ADAPT procedure. Spectacular advantages over ADAPT-VQE are observed for strongly correlated systems including massive savings in circuit depth. Since this compression strategy can also be initialized with accurate Selected-Configuration Interaction (SCI) classical target wave-functions, it paves the way for chemically accurate simulations of larger systems, and strengthens the promise of decisively surpassing classical quantum chemistry through the power of quantum computing.
\end{abstract}

%% file: introductionn.tex
The computational cost of approximating the ground state energy of an $n$-electron molecular system on classical computing architectures typically grows exponentially in $n$. Quantum computers allow for the encoding of the exponentially scaling underlying Hilbert space using only $\mathcal{O}(n)$ qubits, and are therefore likely to outperform classical devices on a range of chemical simulations \cite{aspuru2005simulated, cao2019quantum, mcardle2020quantum}. The Variational Quantum Eigensolver (VQE) is a hybrid quantum-classical algorithm that is considered a very promising candidate for chemical calculations on Noisy Intermediate Scale Quantum (NISQ) devices \cite{peruzzo2014variational, mcclean2016theory}. In this approach, a parameterized wave-function is generated and variationally tuned to minimize the expectation value of the molecular electronic Hamiltonian. A variety of different parameterized wave-functions have been proposed, including the Trotterised Unitary Coupled Cluster (tUCC) ansatz \cite{bartlett1989coupled, romero2018strategies} which consists of a sequence of exponential, unitary operators acting on a judiciously chosen reference state. While the tUCC approach includes electronic correlation and has, in principle, a rather simple quantum circuit structure, the excessive depth of these quantum circuits make them ill-suited for applications in the NISQ regime. This issue has led to the proposal that ansatz wave-functions be constructed through the action of a selective \emph{subset} of possible unitary operators, i.e., only those operators whose inclusion in the ansatz can potentially lead to the largest decrease in the expectation value of the molecular electronic Hamiltonian. In this context, the Adaptive Derivative-Assembled Pseudo-Trotter VQE (ADAPT-VQE) \cite{grimsley2018adapt} has emerged as the gold standard for generating highly accurate and compact ansatz wave-functions. In ADAPT-VQE, the ansatz is grown iteratively by appending a sequence of unitary operators to the reference Hartee-Fock state. At each iteration, the unitary operator to be applied is chosen according to a simple criterion based on the gradient of the expectation value of the Hamiltonian (see the section on technical background and methods for details). 

Assuming that the number of spin-orbitals $N$ being considered is proportional to the number of electrons $n$ in the system, the pool of potential unitary operators in tUCC-based VQEs scales as $\mathcal{O}(N^\ell)$ for $\ell \geq 4$ \footnote{For instance, the pool of operators in tUCCSD scales as $\mathcal{O}(N^4)$, that of tUCCSDT scales as $\mathcal{O}(N^6)$, etc. }. Consequently, conventional VQEs based on the tUCC ansatz require the representation of a product of $\mathcal{O}(N^{\ell})$ unitary operators on quantum circuitry and the optimization of an $\mathcal{O}(N^{\ell})$-dimensional cost-function, both of which are practically impossible using the current generation of NISQ devices. The ADAPT-VQE algorithm attempts to alleviate these problems by avoiding the inclusion of unitary operators in the ansatz wave-function that are not expected to lead to a lowering of the resulting energy. Numerical evidence suggests that ADAPT-VQE is indeed resource-saving and the energy-gradient criterion employed by ADAPT-VQE leads to much more accurate wave-functions than conventional VQE algorithms while preserving moderate circuit depth \cite{grimsley2018adapt,yordanov2021qubit,tang2021qubit}. Thus, while the state-of-the-art k-UpCCGSD algorithm \cite{lee2018generalized}, which the review article \cite{tilly2022variational} considers the most promising fixed-ansatz VQE, is shown to obtain an accuracy of about $10^{-6}$ Hartree for the BeH$_2$ molecule at equilibrium distance at a cost of more than 7000 CNOT gates \cite[Table 1]{xie2022qubit}, ADAPT-VQE achieves a higher accuracy of about $2\times 10^{-8}$ Hartree for the same system using only about 2400 CNOT gates \cite{yordanov2021qubit}. In spite of this comparative advantage, such an energy gradient guided procedure has a tendency to fall into local minima of the energy landscape. Exiting from such minima comes at the expense of adding and optimizing operators through multiple ADAPT iterations \cite{grimsley2022adapt} and leads to over-parameterized wave-functions. In practice, this is associated with an unnecessary increase of the quantum circuit depth required for the representation of the ansatz wave-function coupled to an increasingly difficult classical optimization. This is dramatically revealed in \cite[Supplementary information]{yordanov2021qubit} wherein the basic QEB variant of ADAPT-VQE is applied to the strongly correlated stretched H$_6$ linear chain, and it is shown that more than a thousand CNOT gates are required to construct a chemically accurate ansatz. Given that the current state-of-the-art simulations on physical quantum computers typically involve a maximal circuit depth of less than 100 CNOT gates (see, e.g., the very recent study \cite{zhao2022orbitaloptimized}), it seems unrealistic in the very short term to expect a chemically accurate QPU implementation of ADAPT-VQE for strongly correlated molecules. Let us remark here that while the focus of this article is on hybrid quantum-classical adaptive algorithms in the tradition of ADAPT-VQE, quantum imaginary time evolution approaches have also been recently proposed and shown an improved optimization in the high-dimensional non-convex energy landscape \cite{gomes2021adaptive}.

Our proposed approach for overcoming the challenges of energy plateaus requires modifying the manner in which the ansatz wave-function is constructed. Indeed, rather than constructing an ansatz wave-function through an energy minimisation procedure and potentially encountering local minima, we grow the ansatz wave-function through a process that maximizes its overlap with a -- potentially intermediate -- target wave function that already captures some electronic correlation of the system. We thus use such a target wave-function as a guide to help us build our ansatz in the right direction so as to catch the bulk of electronic correlation. The workflow of this routine is depicted in Figure \ref{fig:workflow}. The resulting overlap-guided ansatz is subsequently used as a high accuracy initialization for an ADAPT-VQE procedure, an algorithm that we refer to as Overlap-ADAPT-VQE. We benchmark and compare the ansatz wave-functions obtained with Overlap-ADAPT-VQE method to standard ADAPT-VQE on a range of small chemical systems with varying levels of correlation. 

%% file: technical_background.tex
\subsubsection*{Qubit Representation of the Molecular Hamiltonian}\label{Sec:2a}
The molecular electronic Hamiltonian with one-body and two-body interactions can be expressed in second-quantization notation as
\begin{equation} \label{eq:2}
	\begin{split}
		H:=\sum_{p,q}h_{pq}a_p^\dagger{}a_q+\sum_{p,q,r,s}h_{pqrs}a_p^\dagger{}a_r^\dagger{}a_sa_q.
	\end{split}
\end{equation}
Here, $p,q,r,$ and $s$ are indices that label the spin-orbitals used to discretize the system, $a_p$ and $a_p^\dagger{}$ are the $p^{\rm th}$ fermionic annihilation and creation operators that satisfy the anti-commutation relations:
\begin{equation} \label{eq:3}
	\left\{a_p,a_q^\dagger{}\right\}:= a_p a_q^\dagger{} +a_q^\dagger{} a_p=\delta_{pq} \quad \text{and}\quad \left\{a_p,a_q\right\}:=a_pa_q + a_qa_p =0,
\end{equation}
with $\delta_{pq}$ representing the classical Kronecker symbol in the frame of operator algebra, and $h_{pq}$ and $h_{pqrs}$ are one-electron and two-electron integrals that can be computed on classical hardware through the expressions
\begin{equation} \label{eq:4}
	\begin{split}
		h_{pq} :=& \int_{\mathbb{R}^3} \Psi_p^*(\bold{x})\left(-\frac{1}{2}\Delta -V_{\rm nuc}\right)\Psi_q(\bold{x})\; d\bold{x}, \\[0.5 cm]
		h_{pqrs} :=& \int_{\mathbb{R}^3} \int_{\mathbb{R}^3} \Psi_p^*(\bold{x})\Psi_r^*(\bold{y})\left(\frac{1}{|\bold{x}-\bold{y}|}\right)\Psi_q(\bold{x})\Psi_s(\bold{y})\; d\bold{x}d\bold{y},
	\end{split}
\end{equation}
where $\Psi_p, \Psi_q, \Psi_r, \Psi_s$ denote spin-orbitals labeled by the indices $p, q, r$, and $s$ respectively.

In order to represent the second-quantized Hamiltonian $H$ on a quantum computer, we use the Jordan-Wigner transform \cite{jordan1993paulische,batista2001generalized} to map the creation and annihilation operators to tensor products involving unitary matrices. To this end, we denote by $\ket{0}_p$ and $\ket{1}_p$ states corresponding to an \emph{empty} and \emph{occupied} spin-orbital $p$ respectively. Using this formalism, the reference Hartree-Fock state for a system having $n$ electrons in $N$ spin-orbitals can be expressed as $\ket{\Psi_{\text{HF}}}:= \ket{1_{0}\ldots 1_{n}0_{n+1}\ldots 0_N}$, and the corresponding fermionic creation and annihilation operators are given by
\begin{equation} \label{eq:jw}
	\begin{split}
		a_p &=\Bigg(\bigotimes_{i=0}^{p-1}Z_i\Bigg) \otimes \frac{X_p+iY_p}{2}  ~=: \Bigg(\bigotimes_{i=0}^{p-1}Z_i \Bigg) \otimes Q_p ,
		\\[0.5cm]
		a_p^\dagger{} &=\Bigg(\bigotimes_{i=0}^{p-1}Z_i\Bigg)  \otimes \frac{X_p-iY_p}{2}~ =: \Bigg(\bigotimes_{i=0}^{p-1}Z_i \Bigg) \otimes Q_p^\dagger{},
	\end{split}
\end{equation}
where $X_p, Y_p, Z_p$ are single qubit Pauli gates applied to qubit $p$ \cite{nielsen00}. Note that in Equation \eqref{eq:jw}, we have introduced the so-called qubit excitation and de-excitation operators $Q_p$ and $Q_p^{\dagger}$ respectively that switch the occupancy of the spin-orbital. These operators will be the subject of further discussion in the sequel. Let us also remark here that the Jordan-Wigner-transformed excitation and de-excitation operators \eqref{eq:jw} respect the anti-commutation relations \eqref{eq:2}. This is simply a consequence of including the tensor product of $Z$-Pauli gates in Equation \eqref{eq:jw} \cite{jordan1993paulische}.



\subsubsection*{The Variational Quantum Eigensolver}
Equipped with the single-qubit Pauli gate representation of the  molecular Hamiltonian $H$, we are now interested in approximating its ground-state eigenvalue. The Variational-Quantum-Eigensolver (VQE) is a hybrid quantum-classical algorithm that couples a classical optimization loop to a subroutine that computes on a quantum computer, the expectation value of the Hamiltonian with respect to a proposed ansatz wave-function. This quantum subroutine involves two fundamental steps:
\begin{enumerate}
	\item The preparation of a trial quantum state (the ansatz wave-function) $\ket{\Psi(\vec{\theta})}$. A variety of different functional forms for the ansatz wave-function have been proposed \cite{kandala2017hardware, romero2018strategies, o2016scalable, liu2019variational} including the aforementioned tUCC ansatz which consists of a sequence of parameterized, exponential fermionic excitation and de-excitation operators acting on a reference state (see below for explicit expressions of these operators).
	
	\item The measurement of the expectation value $\bra{\Psi(\vec{\theta})}H\ket{\Psi(\vec{\theta})}$.
\end{enumerate}

The output of the quantum subroutine is fed into a classical optimization algorithm which calculates the optimal set of parameters $\vec{\theta}_{\rm opt}$ that minimizes the expectation value of the Hamiltonian $H$. The variational principle ensures that the resulting optimized energy is always an upper bound for the exact ground-state energy $E_0$ of $H$, i.e., 
\begin{equation}
	\bra{\Psi(\vec{\theta}_{\rm opt})}H\ket{\Psi(\vec{\theta}_{\rm opt})}\geq E_0.
\end{equation}

The fundamental challenge in implementing the VQE methodology on NISQ devices is thus to construct an ansatz wave-function that can capture the most important contributions to the electronic correlation energy and, at the same time, is capable of being represented on rather shallow quantum circuits. A necessary condition to achieve the latter is that the chosen ansatz wave-function be parameterized with a relatively small number of optimization parameters. Thus, the major computational shortcoming of the popular tUCCSD method-- which otherwise possesses an attractive functional form \cite{romero2018strategies}--  is that its actual implementation on quantum computers requires extremely deep circuits which generate far too much noise on the current generation of NISQ devices \cite{sennane2023calculating}. Indeed, implementing the tUCCSD algorithm on quantum architectures through the Jordan-Wigner mapping \eqref{eq:jw} requires $O(N^3n^2)$ quantum gates  \cite{romero2018strategies} (recall that $N$ is the number of spin-orbitals being considered and $n$ is the number of electrons in the system so that if $N$ is proportional to $n$, then the number of quantum gates required will be of the order of $O(N^5)$). This problem is further exacerbated by the ubiquitous usage of CNOT gates in the construction of quantum circuits for fermionic excitation and de-excitation operators. tUCCSD has been recently extended to triple excitations (tUCCSDT)\cite{UCCSDT} and coupled to both spin and orbital symmetries to reduce the operators count but this latter remains too high for real life QPUs implementation despite a significant increased accuracy over tUCCSD.



\subsubsection*{The ADAPT-VQE Ansatz}
\label{adapt}

The adaptive derivative-assembled pseudo-Trotter variational quantum eigensolver (ADAPT-VQE)\cite{grimsley2018adapt} was designed to overcome the computational shortcomings of the traditional tUCCSD method by proposing an ansatz function that is adaptively grown through an iterative process. ADAPT-VQE is based on the fact \cite{evangelista2019exact} that the full-CI quantum state can be represented by the action of a potentially infinitely long product of only one-body and two-body operators on the reference Hartree-Fock determinant, i.e.,
\begin{equation}
	\ket{\Psi_{\rm FCI}} = \prod_{k}^{\infty}
	\biggl[\prod_{pq}\hat{A}_p^q(\theta^{pq}_k)\prod_{pqrs}\hat{A}_{pq}^{rs}(\theta^{pqrs}_k)\biggr]\ket{\Psi_{\rm HF}}.
\end{equation}
Here, $\hat{A}_p^q(\theta^{pq}_k):= e^{\theta^{pq}_k\hat{\tau}_p^q(k)}$ and $\hat{A}_{pq}^{rs}(k):= e^{\theta^{pqrs}_k\hat{\tau}_{pq}^{rs}}$ where $\hat{\tau}_p^q$ and $\hat{\tau}_{pq}^{rs}$  denote the anti-symmetric operators $\hat{a}_p^q - \hat{a}_q^p$ and $\hat{a}_{pq}^{rs} - \hat{a}_{rs}^{pq}$ and $\theta^{pq}_k$ (resp. $\theta^{pqrs}_k$) is the expansion coefficient of the $k^{\rm th}$ repetition of the operator $\hat{A}_p^q$ (resp. $\hat{A}_{pq}^{rs}$).

The general workflow of the ADAPT-VQE algorithm is as follows:
\begin{enumerate}
	\item On classical hardware, compute one-electron and two-electron integrals, and map the molecular Hamiltonian into a qubit representation. \textbf{On quantum hardware}, boot the qubits to an initial state $\ket{\Psi^{0}} = \ket{\Psi_{\rm{HF}}}$. 
	
	\item Define a pool of parameterized unitary operators that will be used to construct the ansatz. 
	
	\item \textbf{On quantum hardware,} at the $m^{\rm{th}}$ iteration, identify the parameterized unitary operator $\hat{\mathcal{U}}_m(\theta_m)$ whose action on the current ansatz $\ket{\Psi^{m-1}}$ will produce a new wave-function with the largest drop in energy. This identification is done by computing suitable gradients at $\theta_m=0$, the gradients being expressed in terms of commutators involving the molecular Hamiltonian acting on the current ansatz wave-function:
     \begin{equation}
        \frac{\partial}{{\partial \theta_m}} \braket{\Psi^{m-1}|~\hat{\mathcal{U}}_m^\dagger{}(\theta_m)H\hat{\mathcal{U}_m}(\theta_m)|\Psi^{m-1}}\big\vert_{\theta_m=0} = \braket{\Psi^{m-1}|[H,\hat{\mathcal{U}_m}(0)]|\Psi^{m-1}}
    \end{equation}
	
	\item Exit the iterative process if the gradient norm is smaller than some threshold $\epsilon$. Otherwise, append the selected operator to the left of the current ansatz wave-function $\ket{\Psi^{m-1}}$, i.e., define $\ket{\widetilde{\Psi^{m}}}:= \hat{\mathcal{U}_m}(\theta_m) \ket{\Psi^{m-1}}=\hat{\mathcal{U}_m}(\theta_m) \hat{\mathcal{U}}_{m-1}(\theta_{m-1}') \ldots \hat{\mathcal{U}_1}(\theta_1') \ket{\Psi^0}$.
	
	\item \textbf{Hybrid Quantum-Classical VQE:} Optimize all parameters $\theta_m, \theta_{m-1}, \ldots, \theta_1$ in the new ansatz wave-function so as to minimize the expectation value of the molecular Hamiltonian, i.e., solve the optimization problem
    \begin{align}
        \vec{\theta}^{\rm{opt}} :=& (\theta_1', \ldots, \theta_{m-1}', \theta_m')\\ \nonumber
        :=& \underset{\theta_1, \ldots, \theta_{m-1}, \theta_{m}}{\operatorname{argmin}}\braket{\hat{\mathcal{U}_m}(\theta_m)\hat{\mathcal{U}}_{m-1}(\theta_{m-1}) \ldots \hat{\mathcal{U}_1}(\theta_1) \Psi^0|H\hat{\mathcal{U}_m}(\theta_m) \hat{\mathcal{U}}_{m-1}(\theta_{m-1}) \ldots \hat{\mathcal{U}_1}(\theta_1)\Psi^0}
    \end{align}
    and define the new ansatz wave-function $\ket{\Psi^{m}}$ using the newly optimized parameters $\theta_1', \ldots, \theta_m'$, i.e., define $\ket{\Psi^{m}}:= \hat{\mathcal{U}_m}(\theta_m') \hat{\mathcal{U}}_{m-1}(\theta_{m-1}') \ldots \hat{\mathcal{U}_1}(\theta_1') \ket{\Psi^0}$. Let us emphasize that although we also denote the newly optimized parameters at the current $m^{\rm th}$ iteration by $\theta_1',\ldots \theta_m'$, these optimized values are not necessarily the same as those used to define $\ket{\Psi^{m-1}}$ and referenced in Step 4 above.

    \item Return to Step 3 with the updated ansatz $\ket{\Psi^{m}}$. 

\end{enumerate}
   
There are essentially three types of operator pools that are used to construct the ADAPT-VQE ansatz.

\begin{figure}[t]
		\centering
		\begin{equation} \label{staircase}
			\begin{tikzpicture}
				\node[scale=0.9]{
					\begin{quantikz}
						q_k&& \gate{R_x(-\pi/2)}
						& \ctrl{1} & \qw & \qw & \qw & \qw & \qw & \ctrl{1} & \gate{R_x(\pi/2)} & \qw \\
						q_{k-1}&& \qw & \targ{} & \ctrl{1} & \qw & \qw & \qw & \ctrl{1} & \targ{} & \qw & \qw \\
						\vdots &&&& \vdots &&&& \vdots &&& \\
						q_{i+1}&& \qw & \qw & \targ{} & \ctrl{1} & \qw & \ctrl{1} & \targ{} & \qw & \qw & \qw \\
						q_i&& \gate{H} & \qw & \qw & \targ{} & \gate{R_z(\theta)} & \targ{} & \qw & \qw & \gate{H} & \qw \\
				\end{quantikz}};
			\end{tikzpicture}
		\end{equation}
        \begin{equation} \label{eq:sqeb}
			\begin{tikzpicture}
				\node[scale = 1]{
					\begin{quantikz}
						q_p&& \qw & \qw & \gate{R_z(\frac{\pi}{2})} & \ctrl{1} & \gate{R_y(\frac{\theta}{2})}) &\ctrl{1} & \gate{R_y(-\frac{\theta}{2})}& \ctrl{1} &\qw\\
						q_{q}&& \qw & \gate{R_y(-\frac{\pi}{2})} & \gate{R_z(-\frac{\pi}{2})} & \targ{} & \gate{R_z(-\frac{\pi}{2})} & \targ{}& \gate{H}& \targ{} & \qw\\
				\end{quantikz}};
			\end{tikzpicture}
		\end{equation}
		\begin{equation} \label{eq:dqeb}
			\begin{tikzpicture}
				\node[scale=0.5]{
					\begin{quantikz}
						q_s&& \ctrl{1} & \qw & \ctrl{2} & \gate{R_y(\frac{\theta}{8})} & \ctrl{1} & \gate{R_y(-\frac{\theta}{8})} &\ctrl{3} & \gate{R_y(\frac{\theta}{8})} &\ctrl{1} & \gate{R_y(-\frac{\theta}{8})} &\ctrl{2} & \gate{R_y(\frac{\theta}{8})} &\ctrl{1} & \gate{R_y(-\frac{\theta}{8})} &\ctrl{3} & \gate{R_y(\frac{\theta}{8})} &\ctrl{1} & \gate{-R_y(\frac{\theta}{8})} &\ctrl{2} & \gate{R_z(\frac{\pi}{2})} & \qw & \ctrl{1} & \qw\\
						q_{r}&& \targ{} & \gate{X} & \qw & \gate{H} & \targ{}& \qw &\qw &\qw &\targ{}& \qw & \qw &\qw & \targ{} & \qw &\qw & \qw &\targ{} & \gate{H} & \qw & \qw & \gate{X} & \targ{} & \qw\\
						q_{q}&& \ctrl{1} & \qw & \targ{} & \qw &\qw &\qw &\qw &\qw &\qw &\gate{H}&\targ{} & \qw &\qw &\qw &\qw &\qw &\qw &\gate{R_z(-\frac{\pi}{2})} & \targ{} & \gate{R_z(-\frac{\pi}{2})} & \gate{R_y(-\frac{\pi}{2})}& \ctrl{1} & \qw \\
						q_p&& \targ{} & \gate{X}&\qw & \qw &\qw &\gate{H} & \targ{} & \qw &\qw &\qw &\qw &\qw &\qw &\qw & \targ{} & \gate{H} & \qw &\qw &\qw &\qw & \gate{X} & \targ{} & \qw \\
				\end{quantikz}};
			\end{tikzpicture}
		\end{equation}
		\caption{On top, quantum circuit applying the operator $e^{\theta(a_i^\dagger{}a_k)}$ as part of a single fermionic excitation. Double fermionic excitations are carried out using an analogous circuit. In the middle, quantum circuit (\ref{eq:sqeb}) performing a generic single-qubit evolution $U_{pq}^{(\rm{sq})}(\theta)$  and a quantum circuit (\ref{eq:dqeb}) performing a generic double-qubit evolution $U_{pqrs}^{(\rm{dq})}(\theta)$ at the bottom. Note that the terms single-qubit and double-qubit excitations refer to the fact that these operator perform rotations on one pair and two pairs of qubits respectively, not one and two individual qubits.
		\label{fig:qeb_circuits}}
	\end{figure}
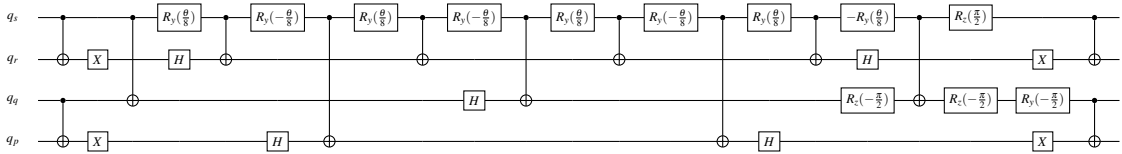

\begin{itemize}
	\item Fermionic-ADAPT-VQE \cite{grimsley2018adapt} uses a pool of spin-complemented pairs of single and double fermionic excitation operators. The quantum circuits performing these unitary operations are of the staircase shape (see Figure \ref{fig:qeb_circuits}, equation  (\ref{staircase})).
	
	\item Qubit-ADAPT-VQE \cite{tang2021qubit} divides the fermionic-ADAPT operators after the Jordan-Wigner mapping and takes the individual Pauli strings as operators of the pool. The quantum circuit for an operator is a single layer of fermionic excitation ``CNOT-staircase'' circuits, similar to the circuit displayed in Figure \ref{fig:qeb_circuits}, equation (\ref{eq:sqeb}).
	
	\item Qubit-Excitation-Based-ADAPT-VQE (QEB-ADAPT-VQE) \cite{yordanov2021qubit} uses a pool of qubit excitation operators. Exponential single-qubit and double-qubit excitation evolutions can be expressed using the qubit creation and annihilation operators $Q_p$ and $Q_p^{\dagger}$ defined through Equation \eqref{eq:jw} as\\
	\begin{equation}
		\begin{split}
			U_{pq}^{\rm (sq)}(\theta)&=\exp{(\theta(Q_p^\dagger{}Q_q - Q_q^\dagger{}Q_p))}\\
			U_{pqrs}^{\rm (dq)}(\theta)&=\exp{(\theta(Q_p^\dagger{}Q_q^\dagger{}Q_rQ_s - Q_r^\dagger{}Q_s^\dagger{}Q_pQ_q))},
		\end{split}
	\end{equation}
	which, after the Jordan-Wigner encoding yields
	\begin{equation}\label{eq:QEB_OP}
		\begin{split}
			U_{pq}^{\rm (sq)}(\theta)=&\exp\Big(-i\frac{\theta}{2}\big(X_qY_p - Y_qX_p\big)\Big)\\
			U_{pqrs}^{\rm (dq)}(\theta)=&\exp\Big(-i\frac{\theta}{8}\big(X_rY_sX_pX_q + Y_rX_sX_pX_q + Y_rY_sY_pX_q + Y_rY_sX_pY_q\\
			&- X_rX_sY_pX_q - X_rX_sX_pY_q - Y_rX_sY_pY_q - X_rY_sY_pY_q\big)\Big),
		\end{split}
	\end{equation}
	with $p, q, r,$ and $s$ denoting, as usual, indices for the spin-orbitals, and we have written (sq) and (dq) as abbreviations for single qubit and double qubit excitation evolutions respectively. The quantum circuits corresponding to the single-qubit and double-qubit excitation operators \cite{yordanov2020efficient} are then given in Figure \ref{fig:qeb_circuits}, equation (\ref{eq:dqeb}).
\end{itemize}
Extensive comparisons between these pools of operators have been carried out by Yordanov et al.\cite{yordanov2021qubit} and numerical evidence suggests that QEB-ADAPT-VQE generates the most computationally tractable ansatz wave-functions. This is primarily due to the fact that qubit excitation circuits can be constructed using much fewer quantum gates than fermionic excitation circuits \cite{yordanov2020efficient} in combination with the observation that qubit excitation evolutions approximate molecular electronic wave-functions with almost the same level of accuracy as fermionic excitation evolutions. For the purpose of this article therefore, we will restrict our attention to operator pools involving qubit excitation evolutions and work in the framework of QEB-ADAPT-VQE.

\subsection*{The Overlap-Guided Adaptative Algorithm (Overlap-ADAPT)}

The numerical evidence presented in the articles \cite{grimsley2018adapt, yordanov2021qubit, grimsley2022adapt} demonstrates that the ADAPT-VQE algorithm is capable of approximating the ground state Full-CI energy to a very high accuracy. Unfortunately, achieving a suitably accurate approximation to the sought-after energy may require a large number of ADAPT iterations which results both in deep quantum circuits that cannot be implemented on the current generation of NISQ devices as well as an increasingly computationally expensive optimization procedure. This problem is particularly apparent in strongly correlated systems for which the ADAPT algorithm frequently encounters energy plateaus\footnote{During these plateaus, a series of new operators are added to the anszatz without meaningfully reducing the energy} \emph{prior} to achieving the classical chemical accuracy threshold of $10^{-3}$ Hartree. Since quantum chemists are primarily interested in numerical results in the regime $10^{-3}$ to $10^{-4}$ Hartree, i.e., slightly more accurate than the chemical accuracy threshold, it is natural to ask if the ADAPT-VQE procedure could be modified so as to avoid these initial energy plateau slowdowns and achieve the required accuracy using an ansatz compact-enough to be implementable on current NISQ devices. 

To make these ideas more precise, let us first introduce for any natural number $p$, the set of all wave-functions that can be represented by the product of exactly $p$ exponential, one-body and two-body qubit excitation evolution operators acting on the Hartree-Fock reference state:
\begin{align}
	W_p := \left\{ \left(\prod_{k=1}^p \exp\left(\theta_k Q_{p_k}Q^{\dagger}_{p_k}\right)\right) \ket{\Psi_{\rm HF}} \colon \theta_k \in \mathbb{R}, ~Q_{p_k}, Q_{p_k}^{\dagger} ~\text{ defined as in Equation \eqref{eq:jw}} \right\}.
\end{align}
Given now an arbitrary electronic wave-function $\ket{\Psi_{\rm ref}}$, we can define the best approximation of $\ket{\Psi_{\rm ref}}$ in the set $W_p$ as
\begin{equation}\label{eq:best_approx}
	\ket{\Psi^*_p} := \underset{ \ket{\Psi} \in W_p}{\text{argmin}} \Big\Vert \ket{\Psi} -\ket{\Psi_{\rm ref}} \Big\Vert,
\end{equation}
where $\Vert \cdot \Vert$ denotes a suitable norm such as the usual $L^2$ or $H^1$ norms on the space of all electronic wave-functions. The $L^2$-norm and the $H^1$-norm can both be computed on either classical computers or on quantum devices, depending on whether the underlying wave-functions are represented classically or on quantum circuitry. The computation of the $L^2$-norm, however, is more direct and we will therefore adopt this choice of norm for the subsequent numerical simulations considered in this study.




Returning now to Equation \eqref{eq:best_approx}, we see that $\ket{\Psi_p^*} $ is the best approximation of an arbitrary target wave-function $\ket{\Psi_{\rm ref}}$ using a product of exactly $p$ exponential qubit excitation evolution operators acting on the Hartree-Fock reference state. The question we are now interested in answering is the following: If we take the \emph{full-CI wave-function} $\ket{\Psi_{\rm FCI}}$ as the target, does the corresponding best approximation $\ket{\Psi_p^{\rm FCI}} $ defined according to \eqref{eq:best_approx} provide a chemically accurate wave-function for small choices of $p$? More precisely, we wish to explore if for small choices of maximal operator count $p$ it holds that
\begin{align}\label{eq:chem_acc}
	\bra{\Psi_p^{\rm FCI}} H\ket{\Psi_p^{\rm FCI}} - \bra{\Psi_{\rm FCI}} H\ket{\Psi_{\rm FCI}} = \bra{\Psi_p^{\rm FCI}} H\ket{\Psi_p^{\rm FCI}}  -E_0 < 10^{-3}\;{\rm Ha}. 
\end{align}
The answer to this question will be a strong indication as to whether there exists an ansatz wave-function that is simultaneously more compact than the ADAPT-VQE ansatz and which can also capture the bulk of the electronic correlation in the system. Let us emphasise that we are specifically interested in understanding whether we can obtain a more compact ansatz wave-function than that produced by ADAPT-VQE at \emph{chemical accuracy} and not at the level of full-CI accuracy.

\begin{figure}[h!]
	\centering
	\includegraphics[width = 0.86\textwidth, trim={0cm, 0cm, 0cm, 4cm},clip=true]{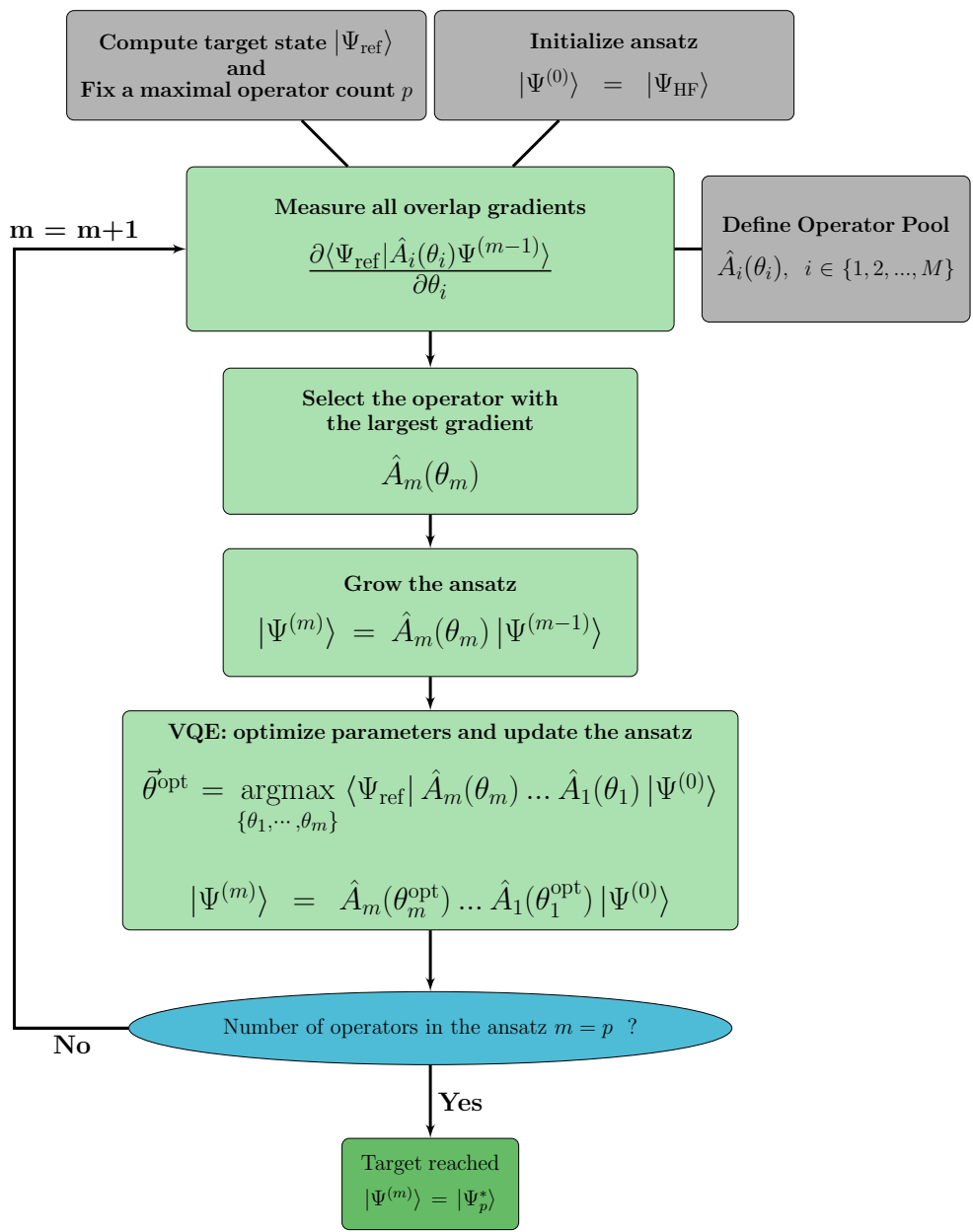}
	\caption{Workflow for Overlap-Guided Adaptative Algorithm (Overlap ADAPT)}
	\label{fig:workflow}
\end{figure}

Unfortunately, answering this question by solving the optimization problem \eqref{eq:best_approx} for an arbitrary target wave-function exactly is not computationally feasible since the size of the set $W_p$ grows exponentially in $p$. Nevertheless, an adaptive, iterative procedure that generates an \emph{approximate} solution to the optimization problem \eqref{eq:best_approx} can  be defined as follows (see also Figure \ref{fig:workflow}). Given a target wave-function $\ket{\Psi_{\rm ref}}$ and a maximal operator count $p$:

\begin{enumerate}
	\item Set the initialisation to the Hartree-Fock reference state, i.e., set $\ket{\Psi^{0}}=\ket{\Psi_{\rm HF}}$.
	
	\item At the $m^{\rm th}$ iteration,  $m\le p$, identify the parametrised exponential qubit excitation evolution operator $\widehat{A}_m(\theta_m)$ whose action on the current ansatz $\ket{\Psi^{m-1}}$ will produce a new wave-function with the largest overlap with respect to the target wave-function. This identification is done by computing the following gradient involving the current ansatz wave-function at $\theta_m=0$:
	\begin{equation}\label{eq:gradient}
		\frac{\partial}{{\partial \theta_m}} \braket{\Psi_{\rm ref}|~\widehat{A}_m(\theta_m)\Psi^{m-1}}\big\vert_{\theta_m=0}.
	\end{equation}

A detailed description of how to compute the gradients given in Equation \eqref{eq:gradient} can be found in the appendix.
 
	\item Append the selected operator to the left of the current ansatz wave-function $\ket{\Psi^{m-1}}$, i.e., define $\ket{\widetilde{\psi^{m}}}:= \widehat{A}_m(\theta_m)\ket{\Psi^{m-1}}= \widehat{A}_m(\theta_m)\widehat{A}_{m-1}(\theta_{m-1}') \ldots \widehat{A}_1(\theta_1')\ket{\Psi^{0}}$.

 \item Optimize all parameters $\theta_m, \theta_{m-1}, \ldots, \theta_1$ in the new ansatz wave-function $\ket{\widetilde{\psi^{m}}}$ so as to maximize its overlap with the target wave-function i.e., solve the optimization problem
	\begin{equation}\label{eq:optimisation}
	\vec{\theta}^{\rm opt}:=(\theta_1',\ldots, \theta_{m-1}', \theta_m' ):=\underset{\theta_1, \ldots, \theta_{m-1}, \theta_{m}} {\text{argmax}}\braket{\Psi_{\rm ref}|\widehat{A}_m(\theta_m)\widehat{A}_{m-1}(\theta_{m-1}) \ldots \widehat{A}_1(\theta_1){\Psi^{0}}}, 
\end{equation}
and define the new ansatz wave-function $\ket{\Psi^{m}}$ using the newly optimized parameters $\theta_1', \ldots, \theta_m'$, i.e., define $\ket{\Psi^{m}}:= \widehat{A}_m(\theta_m')\widehat{A}_{m-1}(\theta_{m-1}') \ldots \widehat{A}_1(\theta_1')\ket{\Psi^{0}}$. Let us emphasize that although we also denote the newly optimized parameters at the current $m^{\rm th}$ iteration by $\theta_1',\ldots \theta_m'$, these optimized values are not necessarily the same as those used to define $\ket{\Psi^{m-1}}$ and referenced in Step 3 above.


	\item If the total number of operators in the updated ansatz is equal to $p$, exit the iterative process. Otherwise go to Step 2 with the updated ansatz wave-function.
\end{enumerate}

We refer to this adaptive procedure as the Overlap-ADAPT-VQE. Let us emphasise here that rather than fixing a maximal operator count, we may employ some other convergence criteria such as the magnitude of the overlap or the magnitude of the gradient vectors as in the original ADAPT-VQE. Moreover, depending on whether the target wavefunction is in a quantum or a classical representation, the gradient screening and the overlap measurements can be performed using either a quantum or a classical device. In particular, if the targeted wave-function is classically computed, then no additional quantum resources are required or measurements are required to compute the overlaps.

We are now interested in applying the Overlap-ADAPT procedure to the reference full-CI wave-functions of some simple, yet strongly correlated molecular systems in an effort to understand the compactness of the wave-function generated by QEB-ADAPT-VQE in the chemical accuracy regime. To do so, we will compute the energy of the Overlap-ADAPT approximation of the target full-CI wave-functions of a stretched BeH$_2$ molecule and a stretched linear H$_6$ chain in a minimal basis set as a function of the number of optimisation parameters, and plot this energy in comparison to the energy obtained using QEB-ADAPT-VQE.

\begin{figure}[h!]
	\centering
	\begin{subfigure}{0.495\textwidth}
		\centering
		\includegraphics[width=\textwidth, trim={0cm, 0cm, 0cm, 0cm},clip=true]{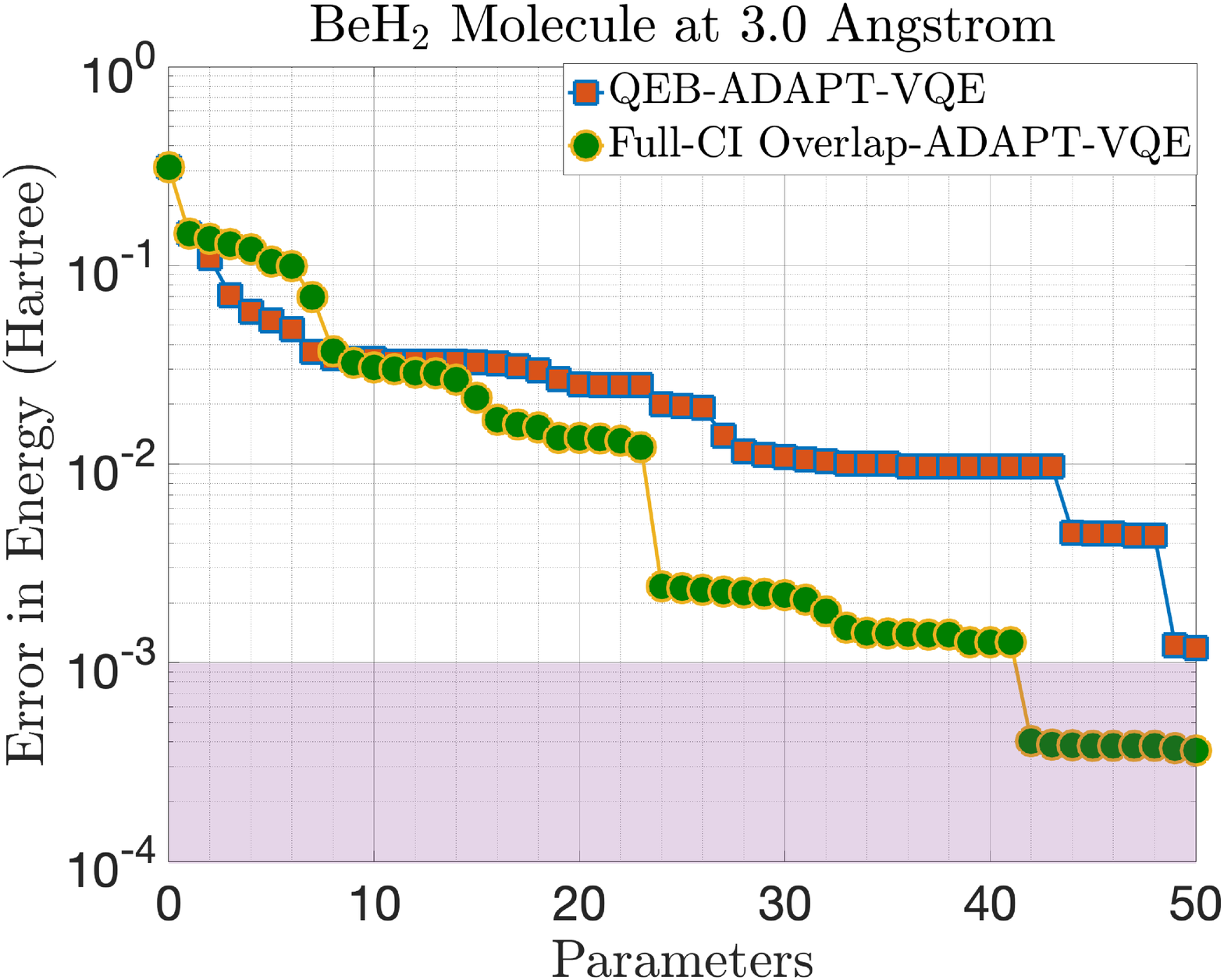} 
	\end{subfigure}\hfill
	\begin{subfigure}{0.495\textwidth}
		\centering
		\includegraphics[width=\textwidth, trim={0cm, 0cm, 0cm, 0cm},clip=true]{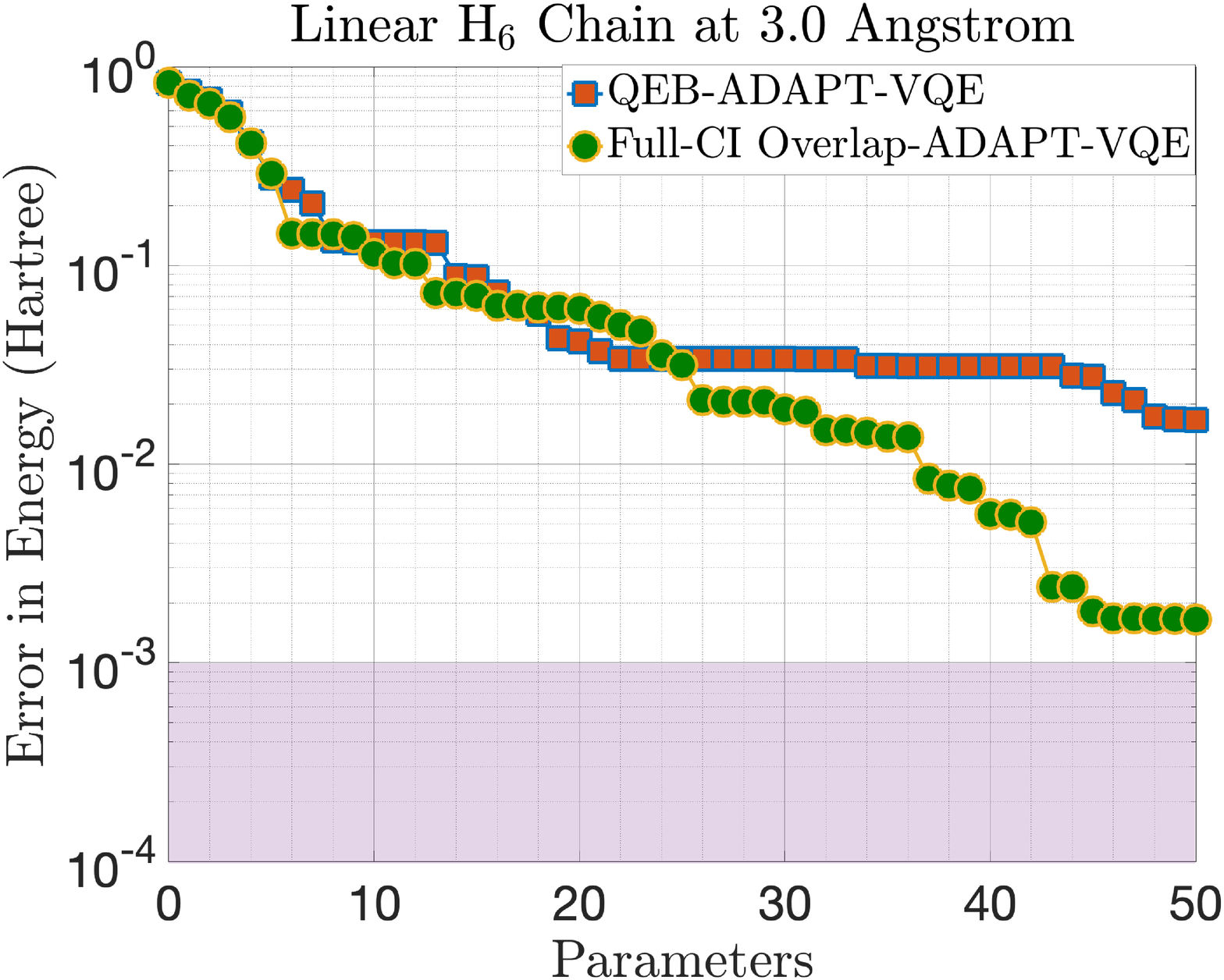} 
	\end{subfigure}
	\caption{Comparison of the Full-CI Overlap Guided ADAPT-VQE and ADAPT-VQE for the ground state energy of a stretched BeH$_2$ molecule and a stretched linear H$_6$ chain with an interatomic distance of 3 Angstrom for both. The plots represent the energy convergence as a function of the number of parameters in the ansatz. The pink area indicates chemical accuracy at $10^{-3}$ Hartree.}
	\label{fig:FCI}
\end{figure}

The resulting energy plots, which are displayed in Figure \ref{fig:FCI}, clearly show that the overlap-guided adaptive procedure is able to avoid the initial energy plateaus afflicting the ADAPT procedure that prevent the attainment of chemical accuracy in a small number of iterative steps.  These results strongly suggest the potential for creating a more condensed ansatz wave-function than that generated by ADAPT-VQE which can sidestep the issue of early energy plateaus. 



\begin{figure}[h!]
	\centering
	\begin{subfigure}{0.495\textwidth}
		\centering
		\includegraphics[width=\textwidth, trim={0cm, 0cm, 0cm, 0cm},clip=true]{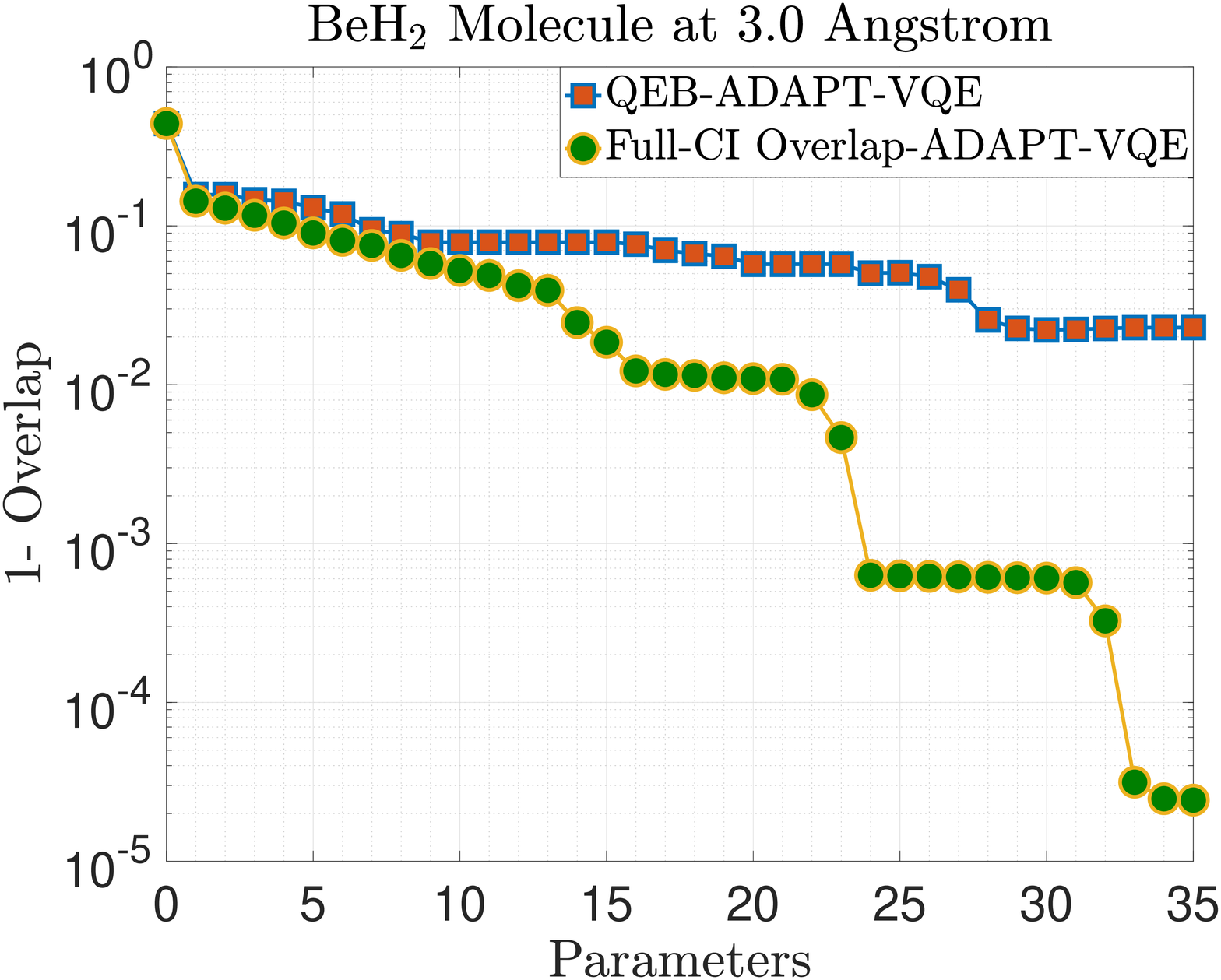} 
	\end{subfigure}\hfill
	\begin{subfigure}{0.495\textwidth}
		\centering
		\includegraphics[width=\textwidth, trim={0cm, 0cm, 0cm, 0cm},clip=true]{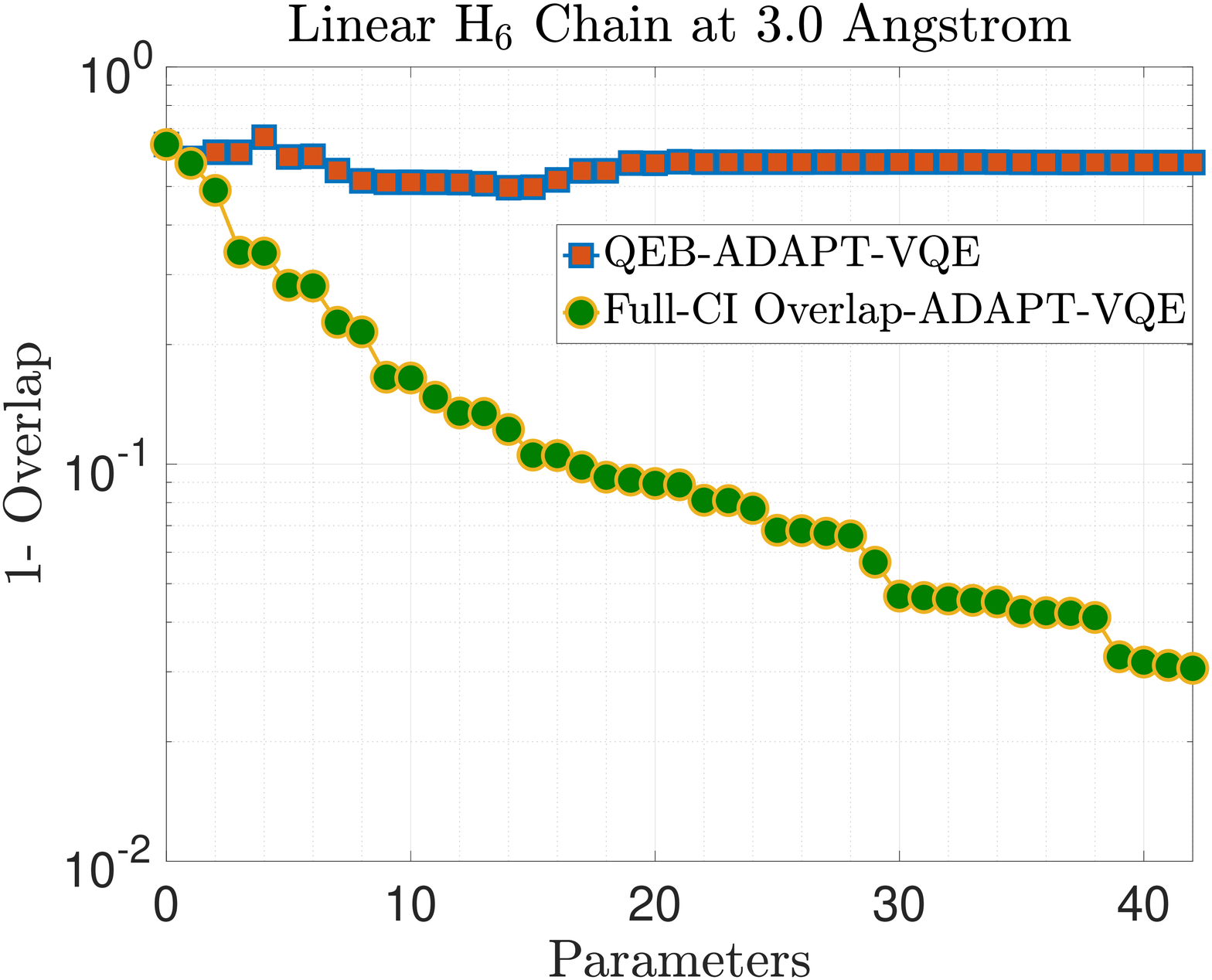} 
	\end{subfigure}
	\caption{Comparison of the Full-CI Overlap Guided ADAPT-VQE and ADAPT-VQE for maximising the overlap with the full-CI wave-function of a stretched BeH$_2$ molecule and a stretched linear H$_6$ chain with an interatomic distance of 3 Angstrom for both. The plots represent the infidelity between the ansatz and the full-CI wave-function, calculated as one minus the overlap, as a function of the number of parameters in the ansatz.}
	\label{fig:FCI_overlap}
\end{figure}

Before proceeding, let us point out that a key metric for evaluating the efficiency of the Overlap-ADAPT algorithm is to compute the overlap between the ansatz wave-function and the full-CI wave-function over the course of several algorithm iterations. Consequently, for the stretched BeH$_2$ and stretched linear H$_6$ chain considered above, we plot the overlap convergence with respect to the full-CI wave-function in Figure \ref{fig:FCI_overlap}. It is readily seen that the Overlap-ADAPT procedure targeted at the full-CI wave-function far outperforms the original ADAPT-VQE, achieving a notably higher overlap with the full-CI wave-function for both a stretched BeH$_2$ molecule and a stretched linear H$_6$ chain. In particular, for the H$_6$ system, while ADAPT-VQE reaches a plateau and stalls its progress, the Overlap-ADAPT procedure smoothly advances without interruption.

Of course the Overlap-ADAPT-VQE targeted at a full-CI wave-function does not define a practical VQE since the full-CI ground state energy is precisely the quantity we wish to approximate. A practical VQE based on orbital overlap optimization can, however, be developed by replacing the targeted full-CI wave-function with a tractable high accuracy approximation thereof and using the resulting overlap-guided ansatz wave-function as a high accuracy initialisation for a new ADAPT-VQE procedure. The targeted ``computable'' wave-function in this situation can be completely general, i.e., it can be the output of any existing numerical algorithm, whether classical or quantum.

The goal of the subsequent sections is to showcase the efficacy of this Overlap-ADAPT algorithm at obtaining chemically accurate results using a minimal number of optimisation parameters. Such findings are important for practical uses of quantum computing for quantum chemistry since, as we have already stated, real-life chemists are interested in reaching convergence in energies corresponding to the so-called chemical accuracy, i.e. $10^{-3}$ to $10^{-4}$ Hartree. Our results can therefore introduce a practical route for compactyfing the ADAPT-VQE operator counts using the Overlap-ADAPT-VQE within this accuracy regime.

%% file: results2.tex
\subsection*{Setting of Numerical Simulations}

The classical numerical simulations reported in this section have been carried out with an in-house code, using Openfermion-PySCF module \cite{PySCF} for integral computations and OpenFermion \cite{mcclean2020openfermion} for second quantization and the Jordan-Wigner mapping. All calculations are performed within the minimal STO-3G basis set\cite{STO3G} without considering frozen orbitals unless otherwise specified. Note that the number of qubits that a simulation requires is equal to the number of spin orbitals of a system, which therefore limits the quality of the single-particle basis and the size of the system that can be simulated. All optimization routines use the BFGS algorithm implemented on the SCIPY Python module \cite{2020SciPy-NMeth}. We use a pool of non spin-complemented restricted single- and double-qubit excitations evolutions. By 'restricted', we mean that we consider only excitations from occupied orbitals to virtual orbitals with respect to the Hartree-Fock determinant. Using fewer operators in the pool makes the gradient screening process faster and easier to handle from a computational point of view \cite{yordanov2021qubit}. To ensure a fair comparison, this same operator pool is used for both the overlap-guided Ansatz and ADAPT-VQE. 

To anticipate applications on noisy quantum machines of such adaptive algorithms, there are essentially two constraints to respect:
\begin{itemize}
\item The circuit depth should be kept as shallow as possible so as to reduce the effect of decoherence in NISQ devices. In the current context, the circuit depth corresponds to the number of gates used to construct our wave-function ansatz.
\item The number of measurements an NISQ device can undertake is very limited. On the other hand, the ADAPT-VQE algorithm requires a large number of measurements both in the form of gradient evaluations at the beginning of each iteration and during the VQE optimization step of the ansatz wave-function. The optimization step in particular often requires an excessive number of measurements since the cost function is both high dimensional and noisy. Consequently, the optimization of the ansatz wave-function is simply intractable with a limited number of evaluations thus preventing practical application of ADAPT-VQE on current quantum devices.
\end{itemize}

In order to implement such adaptive algorithms on the current generation of NISQ devices therefore, we must minimise both the circuit depth and the number of evaluations. Indeed, as the depth of a circuit increases, the noise level also increases, which results in a greater number of samples being required for accurate measurement of the Hamiltonian expectation values. In ADAPT-VQE, each operator added to the ansatz corresponds to an additional layer of quantum gates in the circuit and an additional parameter in the ansatz. Consequently, to address both the circuit depth and the number of evaluations constraints, we will evaluate the energy convergence as a function of the number of operators present in the ansatz.

\subsection*{Application of Overlap-ADAPT-VQE for Compactification of ADAPT-VQE Ansatzë}

As a first test of its effectiveness, we apply the overlap-guided adaptive algorithm to a target wave-function provided by an existing QEB-ADAPT-VQE procedure and then use the result as a high-accuracy initialisation for a new QEB-ADAPT-VQE procedure. Essentially, this first set of numerical experiments is meant to model the situation where we have a strong constraint on the circuit depth (represented by the number of optimisation parameter in the ansatz wave-function), and we wish to see if it is possible to use the Overlap-ADAPT-VQE procedure to compactify the ADAPT-VQE ansatz thereby obtaining a higher accuracy wave-function that respects the constraint on the circuit depth.

\begin{figure}[h!]
	\centering
	\begin{subfigure}{0.495\textwidth}
		\centering
		\includegraphics[width=\textwidth, trim={0cm, 0cm, 0cm, 0cm},clip=true]{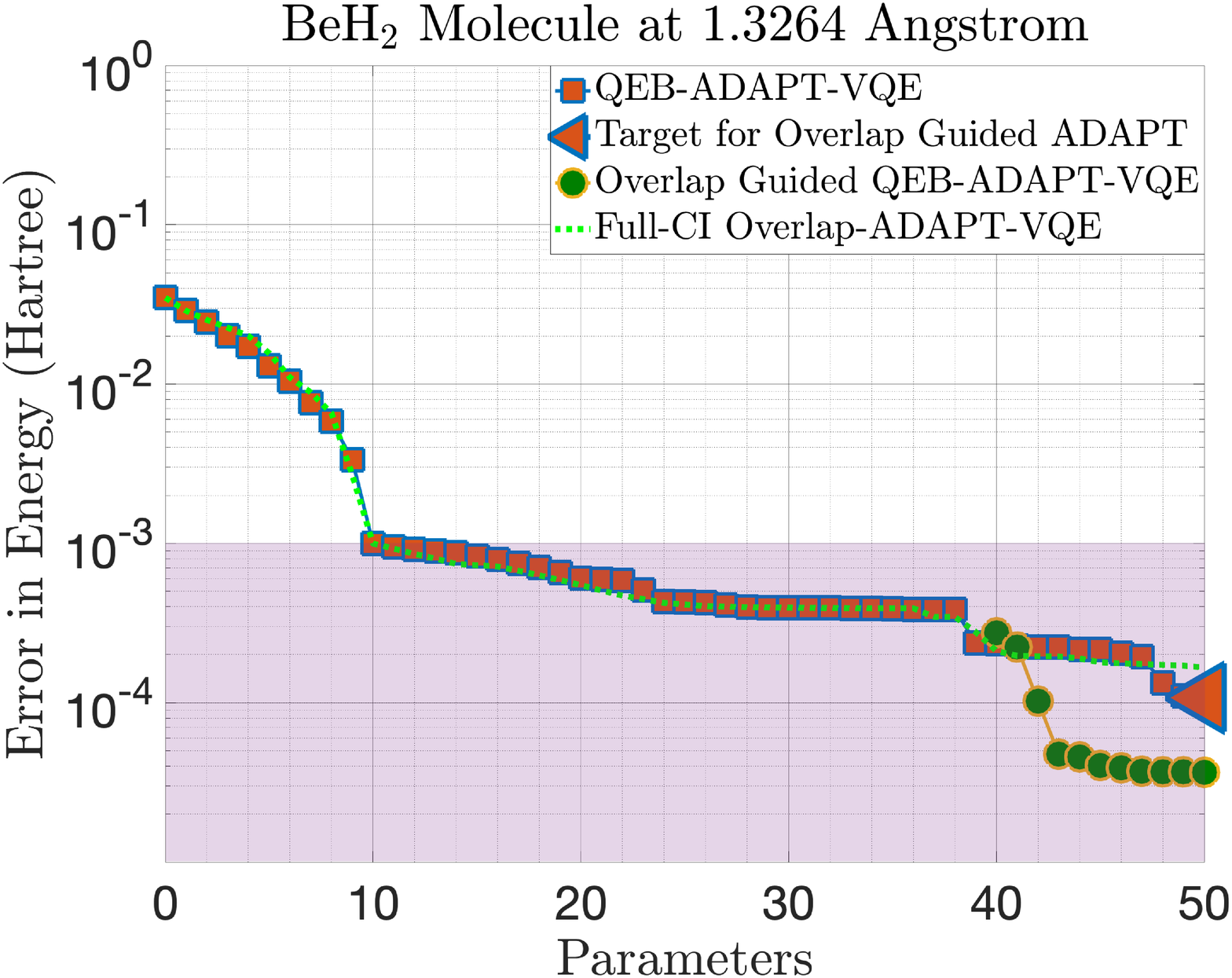} 
	\end{subfigure}\hfill
	\begin{subfigure}{0.495\textwidth}
		\centering
		\includegraphics[width=\textwidth, trim={0cm, 0cm, 0cm, 0cm},clip=true]{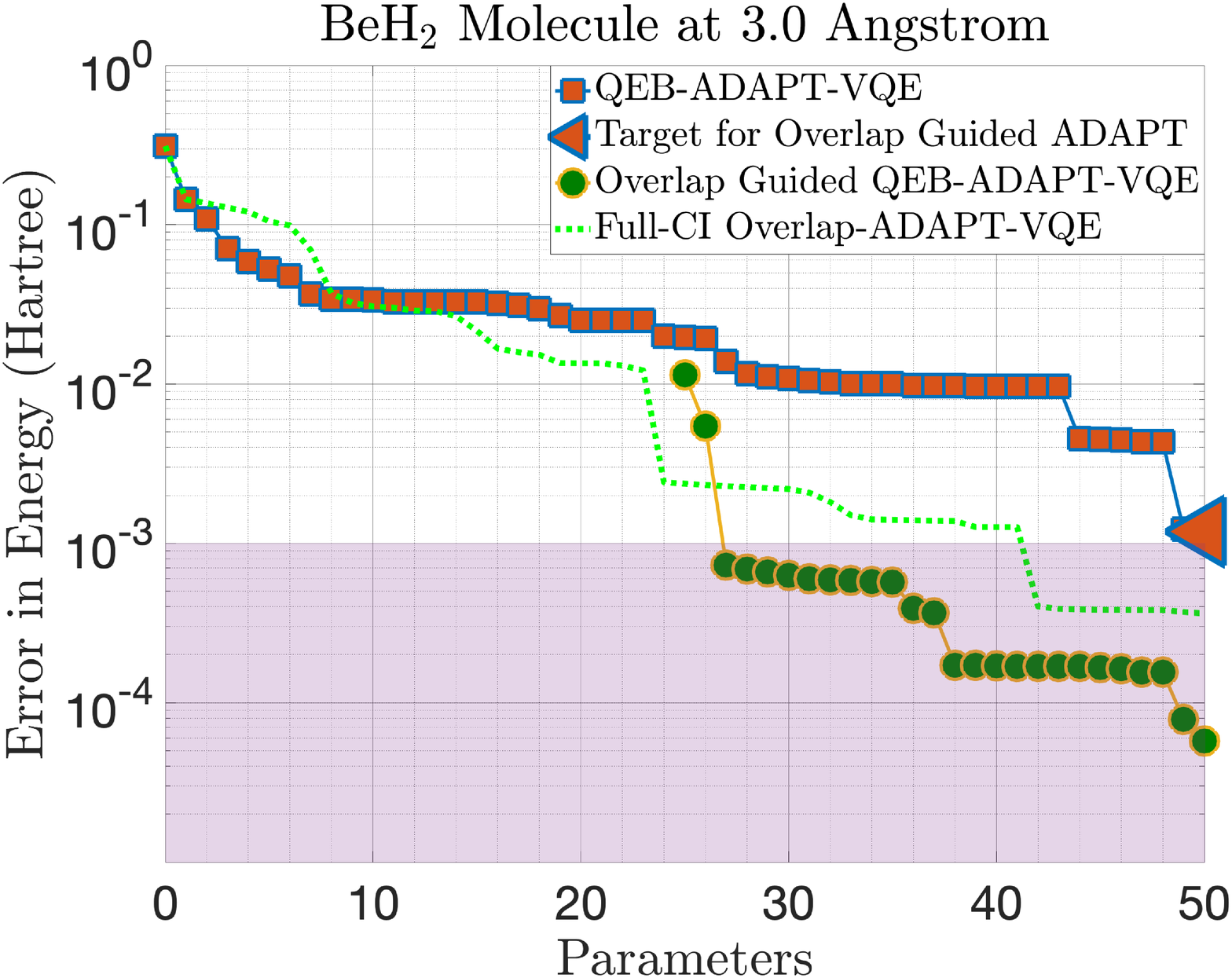} 
	\end{subfigure}
	\caption{Comparison of the Overlap-ADAPT-VQE and ADAPT-VQE for the ground state energy of a BeH$_2$ molecule at equilibrium and stretched geometries. The plot represents the energy convergence as a function of the number of parameters in the ansatz. The left-pointing triangles denote the target wave-functions used for subsequent Overlap-ADAPT procedures. For simplicity, we do not plot the entire Overlap-ADAPT curve, rather only the portion corresponding to the energy minimisation using a classical ADAPT-VQE procedure. Thus, in the case of the left figure, the overlap maximisation portion of Overlap Guided QEB-ADAPT-VQE lasts until parameter 40 at which point the energy minimisation portion is initiated. The green dotted line corresponds to an FCI-Overlap-ADAPT-VQE procedure which is plotted as a reference. Note that at equilibrium distance (the left figure), the QEB-ADAPT-VQE curve and the FCI-Overlap-ADAPT-VQE nearly coincide whereas for the stretched molecule (the right figure) the FCI-Overlap-ADAPT-VQE curve is noticeably lower. The pink area indicates chemical accuracy at $10^{-3}$ Hartree.}
	\label{fig:02a}
\end{figure}

We compute the ground state energy of the benchmark Beryllium Hydride (BeH$_2$) molecule considered in the original ADAPT-VQE articles \cite{grimsley2018adapt}. We consider the BeH$_2$ molecule both at its equilibrium geometry (bond length of 1.3264 Angstrom) as well as at a stretched geometry (bond length of 3.0 Angstrom), which is meant to model a more strongly correlated system. Our results are depicted in Figure \ref{fig:02a}.

\begin{figure}[h!]
	\centering
	\begin{subfigure}{0.495\textwidth}
		\centering
		\includegraphics[width=\textwidth, trim={0cm, 0cm, 0cm, 0cm},clip=true]{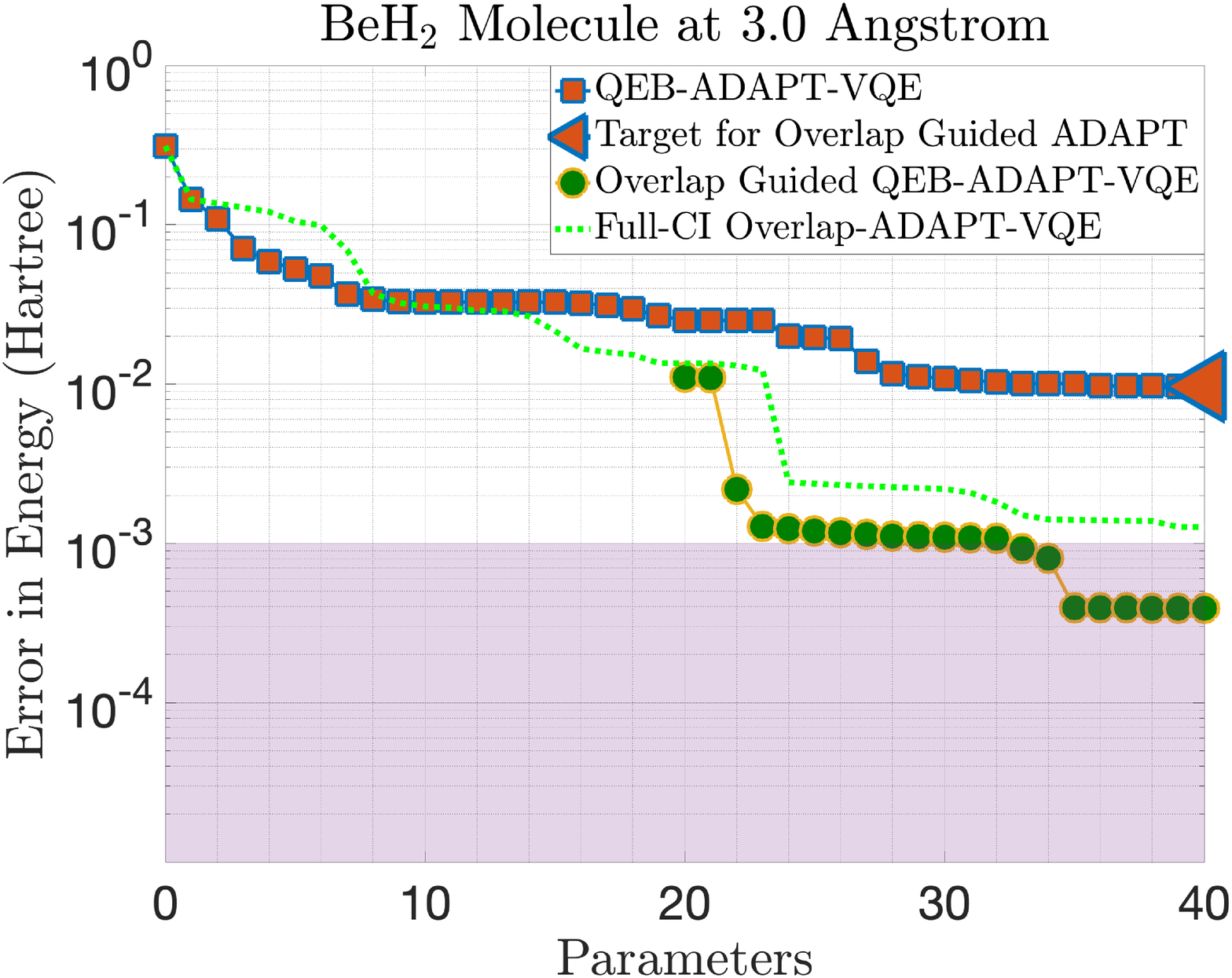} 
	\end{subfigure}\hfill
	\begin{subfigure}{0.495\textwidth}
		\centering
  		\includegraphics[width=\textwidth, trim={0cm, 0cm, 0cm, 0cm},clip=true]{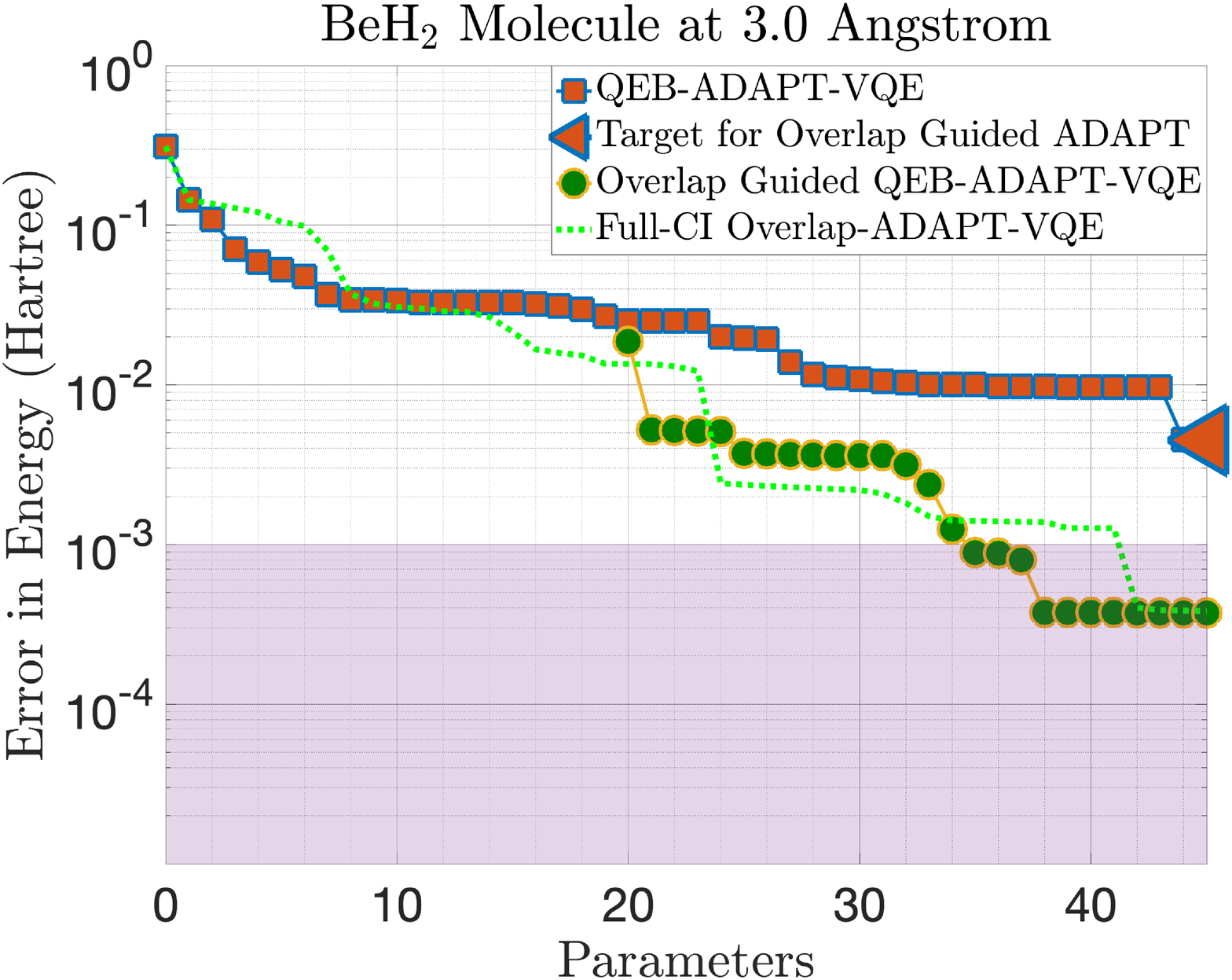} 
	\end{subfigure}
	\caption{Comparison of the Overlap-ADAPT-VQE and ADAPT-VQE for the ground state energy of a stretched BeH$_2$ molecule with a maximal operator count of 40 and 45. The plot represents the energy convergence as a function of the number of parameters in the ansatz. The left-pointing triangles denote the target wave-functions used for a subsequent Overlap-ADAPT procedure. For simplicity, we do not plot the entire Overlap-ADAPT curve, rather only the portion corresponding to the energy minimisation using a classical ADAPT-VQE procedure. The green dotted line corresponds to an FCI-Overlap-ADAPT-VQE procedure which is plotted as a reference. The pink area indicates chemical accuracy at $10^{-3}$ Hartree.}
	\label{fig:02b}
\end{figure}

The numerical results indicate that the Overlap-ADAPT-VQE can indeed compactify the QEB-ADAPT-VQE ansatz wave-function and using the output as an initialization for a new QEB-ADAPT-VQE yields a much more accurate wave-function. Under the constraint of a maximal operator count of 50, the overlap-guided procedure improves the final accuracy of the computed BeH$_2$ ground state energy at equilibrium and stretched geometries by a factor of 3 and 10 respectively. Note that the improvement in accuracy is much higher in the case of the stretched BeH$_2$ molecule which exhibits strong correlation, and this suggests that the comparative advantage of the overlap-guided adaptive algorithm over a pure ADAPT-VQE procedure will be more conspicuous for strongly correlated molecules-- systems for which the ADAPT-VQE algorithm struggles to compute the ground state energy. Thus, in the case of the BeH$_2$ molecule for instance, we are able to achieve chemical accuracy using only a 34 operator-ansatz wave-function whereas the QEB-ADAPT-VQE algorithm requires more than 50. Numerical simulations for stretched BeH$_2$ using a lower maximal operator count of 40 and 45 are displayed in Figure \ref{fig:02b} and show similar improvements in the final accuracy of the ansatz wave-function, although the advantage decreases as the maximal operator count decreases.

\vspace{-1mm}

\vspace{-4mm}

\begin{figure}[H]
	\centering
	\begin{subfigure}{0.495\textwidth}
		\centering
		\includegraphics[width=\textwidth, trim={0cm, 0cm, 0cm, 0cm},clip=true]{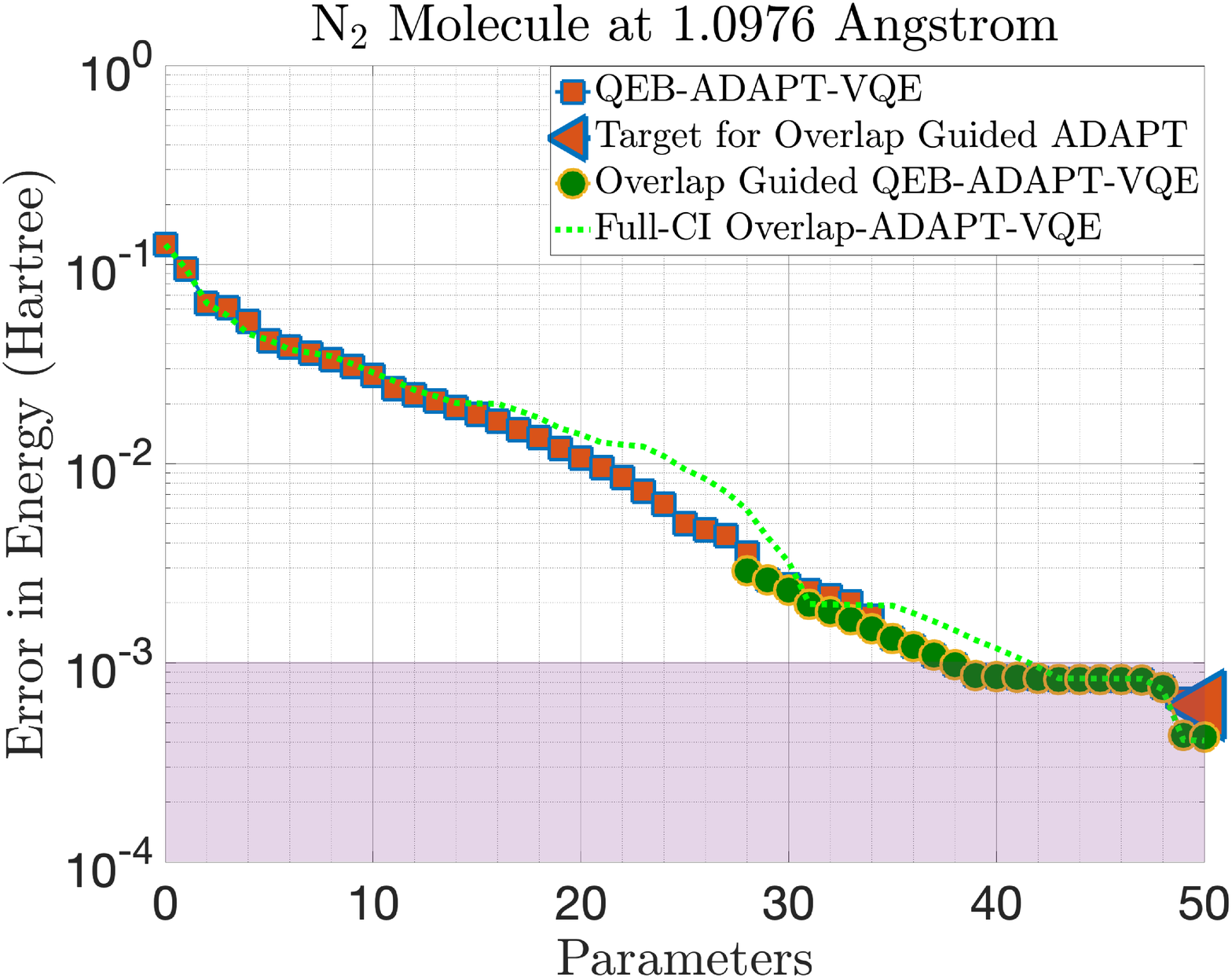} 
	\end{subfigure}\hfill
	\begin{subfigure}{0.495\textwidth}
		\centering
		\includegraphics[width=\textwidth, trim={0cm, 0cm, 0cm, 0cm},clip=true]{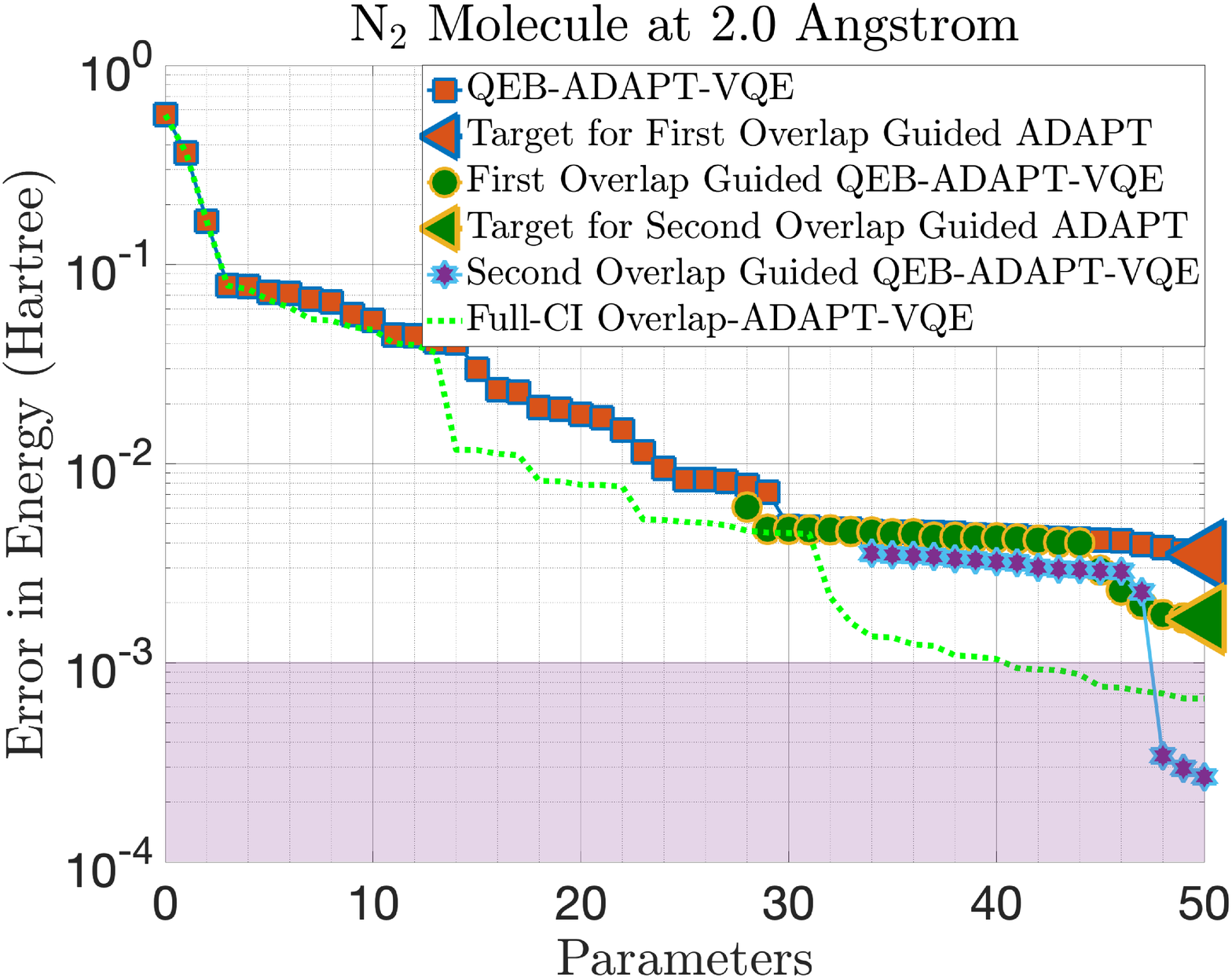} 
	\end{subfigure}
	\caption{Comparison of the Overlap-ADAPT-VQE and ADAPT-VQE for the ground state energy of an N$_2$ molecule at equilibrium and stretched geometries. The plots represent the energy convergence as a function of the number of parameters in the ansatz. The right-pointing triangles denotes the start of an ADAPT-VQE procedure. The left-pointing triangles denote the target wave-functions used for a subsequent Overlap-ADAPT procedure. For simplicity, we do not plot the entire Overlap-ADAPT curve, rather only the portion corresponding to the energy minimisation using a classical ADAPT-VQE procedure. The green dotted line corresponds to an FCI-Overlap-ADAPT-VQE procedure which is plotted simply as a reference. The pink area indicates chemical accuracy at $10^{-3}$ Hartree.}
	\label{fig:N2}
\end{figure}

A further test of the Overlap-ADAPT-VQE applied to a target QEB-ADAPT-VQE wave-function is carried out for the diatomic Nitrogen (N$_2$) molecule at equilibrium and stretched geometries. Although the minimal basis set for N$_2$ is quite large, a tractable computation can be carried out using an active space approach where the eight core electrons of the N$_2$ molecule are frozen and the ground state energy of the system is computed using the resulting frozen core effective Hamiltonian an approach commonly referred as CAS(6,6). As shown in Figure \ref{fig:N2}, we see that the Overlap-ADAPT procedure does not further compactify the QEB-ADAPT-VQE wave-function at equilibrium, the final accuracy of the Overlap-QEB-ADAPT-VQE being only slightly higher than that of the classical QEB-ADAPT-VQE procedure. Nevertheless, by applying the Overlap-ADAPT-VQE procedure \emph{twice}, i.e., taking a QEB-ADAPT-VQE wave-function as the first target, performing an Overlap-ADAPT-VQE procedure, and then taking the resulting wave-function as the target for an additional Overlap-ADAPT-VQE procedure yields a huge gain in accuracy for the stretched geometry. Indeed, the Overlap-QEB-ADAPT-VQE energy is nearly an order of magnitude more accurate than the classical QEB-ADAPT-VQE energy.


Let us remark that as a rule of thumb, for all these simulations, the Overlap-ADAPT algorithm is used to construct an approximate wave-function using a number of operators equal to about 40\%-50\% of the maximal operator count. If the maximal operator count is more flexible, then as a general rule we observe that the ADAPT-VQE ansatz taken immediately after the ADAPT process has exited an energy plateau, serves as an effective choice of target wave-function for an overlap-guided adaptive procedure, i.e., the Overlap-ADAPT-VQE can produce a more compact wave-function with comparable energy to that of the target ADAPT wave-function. On the other hand, taking ADAPT-VQE ansatz wave-function from the middle of an energy plateau as the overlap-guided target seems to be a less effective strategy.



\subsection*{Application of Overlap-ADAPT-VQE to Classically Computed Wave-Functions}

The stretched linear H$_6$ chain is a molecular system that exhibits a high degree of electronic correlation. The complex electronic structure creates a rough energy landscape with many local minima, making the finding of the global energy minimum difficult. This system has already been extensively studied \cite{yordanov2021qubit} and it was shown that achieving chemical accuracy with ADAPT-VQE method required constructing an ansatz wave-function with more than 150 operators from a pool of either generalized fermionic or generalized qubit-excitations. Clearly, resources of this kind are far from being accessible on current NISQ devices, and it is therefore necessary to develop adaptive methods for simulating systems using a much smaller operator count. Until now, the most extensive VQE experiments have typically encompassed around 10 operators while accumulating an error of at least 0.1 Hartree \cite{zhao2022orbital, sennane2023calculating}.  Unfortunately, the ADAPT-VQE ansatz wave-function, presumably not constructed with a satisfactory choice of qubit excitation evolution operators prior to an unreachable number of iterations, cannot be used as the target of the overlap-guided adaptive algorithm as in the previous subsection. Instead, we propose the use of an intermediate, classically computed, multi-configuration wave-function as the overlap-guided target. This approach has the consequent advantage of not costing additional quantum resources. Particularly well-suited choices which fit in the framework of adaptive methods are provided by the so-called Selected-CI (SCI) methods.

\subsubsection*{Combining Classical Selected-CI Approaches and Quantum Computing}
The key idea of SCI methods is to build a compact representation of the reference wave-function by selecting \emph{on-the-fly} the most relevant Slater determinants thanks to an importance criterion based on perturbation theory (PT).
Thanks to this clever selection of the Slater determinants, the variational energy of the reference wave function converges rapidly towards the full-CI energy. Although the recent revival of SCI approaches 
\cite{GinSceCaf-CJC-13,GinSceCaf-JCP-15,hbci,SchEva-JCTC-17,GinTewGarAla-JCTC-18,LooSceBloGarCafJac-JCTC-18,LooBogSceCafJac-JCTC-19,QP2,ZhaLieuHof-JCTC-21} has significantly pushed further the size limit of systems for which near full-CI quality energies can be obtained (typically a few tens of correlated electrons in about two hundreds of orbitals \cite{benzene_bench,LooDamSce-JCP-20}), the scaling of SCI methods is intrinsically exponential in the number of correlated electrons and orbitals. 

The reason for this exponential scaling is directly linked to the linear parametrization of the sought-after wave-function in terms of Slater determinants, which implies that the intrinsic exponential structure of the wave function must be built explicitly by adding more and more determinants to the reference wave function. 
This necessarily leads to size consistency errors which manifest through an underestimation of the coefficients of the reference and perturbative wave functions and therefore of the correlation energy. 
Because the size consistency errors grow with the total (absolute) value of the correlation energy, SCI methods struggle more and more as the number of correlated electrons increases and/or the strength of correlation increases. Recently, attempts to cure this problem have been proposed with a selection of the individual 
excitation operators\cite{XuUejTen-JPCL-20,GurDeuShePie-JCP-21} in a single-reference CC approach.

To overcome these limitations of SCI approaches, an alternative idea is to combine the robust and linear parametrization of SCI with the intrinsic exponential parametrization of the ansatz used in QC computation to take advantage of both worlds: 
\begin{enumerate}
    \item  While reaching chemical accuracy in SCI methods is a struggle in the strong correlation regime, obtaining a compact and robust representation of the bulk of correlation effects is an easy task thanks to the smart selection of Slater determinants and the simplicity of the linear parametrization;

\item Use this compact SCI wave-function as the target of the overlap-guided adaptive algorithm so as to obtain an intermediate wave-function represented in terms of qubit excitation evolution operators acting on the Hartree-Fock reference state;

\item Use the intermediate wave-function as a high accuracy initialization of a new QEB-ADAPT-VQE procedure.
\end{enumerate}

For the purpose of this study, we choose to employ the so-called CI pertubatively selected iteratively (CIPSI) algorithm implemented in QP2\cite{QP2} to generate the required SCI wave-function. Before proceeding to the application of this algorithm to the linear H$_6$ chain, we provide a brief recap of the CIPSI methodology.

\subsubsection*{The CIPSI algorithm in a nutshell}

The CIPSI algorithm, which was originally introduced in the late seventies\cite{HurMalRan-JCP-73,three_class_CIPSI}, is the archetype of SCI approaches: it approximates the FCI wave function through an iterative selected CI procedure, and the FCI energy through a second-order  multi-reference perturbation theory (in this case, with an Epstein--Nesbet\cite{epstein,nesbet} partition). 

The CIPSI energy is defined as
\begin{equation}
  E_\mathrm{CIPSI} := E_\text{v} + E^{(2)}.
\end{equation}
Here, $E_\text{v}$ is the variational energy given by
\begin{equation}
  E_\text{v} := \min_{\{ c_{\rm I}\}} \frac{\elemm{\Psi^{(0)}}{\hat{H}}{\Psi^{(0)}}}{\ovrlp{\Psi^{(0)}}{\Psi^{(0)}}},
\end{equation}
where the reference wave function $\ket{\Psi^{(0)}} = \sum_{{\rm I}\,\in\,\mathcal{R}} \,\,c_{\rm I} \,\,\ket{\rm I}$ is expanded in Slater determinants~$\ket{\rm I}$ within the CI reference space $\mathcal{R}$, and $E^{(2)}$ is the second-order energy correction defined as
\begin{equation}
  E^{(2)} := \sum_{\kappa} \frac{|\elemm{\Psi^{(0)}}{\hat{H}}{\kappa}|^2}{E_\text{v} - \elemm{\kappa}{H}{\kappa}} = \sum_{\kappa} \,\, e_{\kappa}^{(2)},
\end{equation}
where $\kappa$ denotes a determinant outside the reference space $\mathcal{R}$.  

The CIPSI energy is systematically refined by doubling the size of the CI reference space at each iteration, selecting
the determinants $\kappa$ with the largest $\vert e_{\kappa}^{(2)} \vert$. The calculations are stopped when a target value of $E^{(2)}$ is reached.

\subsubsection*{CIPSI-Overlap-ADAPT Numerical results}
We performed CIPSI calculations through the open-source quantum chemistry environment Quantum Package \cite{QP2}for the different molecular systems. As mentioned previously, the CIPSI wavefunction is used as a target for the overlap-guided adaptive algorithm and is therefore not required to be very accurate. In particular, all CIPSI wave-functions employed in this study have error much larger than $10^{-3}$ Hartree, i.e., they are \emph{not} chemically accurate. In the remainder of this section, we compare the energy convergence of the QEB-ADAPT-VQE algorithm starting from an intermediate wave-function obtained by applying the overlap-guided algorithm to a CIPSI wave-function with the traditional QEB-ADAPT-VQE procedure that initializes from a simple Hartree-Fock ansatz. As a rule of thumb, for all these simulations, the Overlap-ADAPT-VQE is used to construct an approximate wave-function with energy comparable to that of the targeted CIPSI wave-function before initiating the subsequent QEB-ADAPT-VQE procedure.

\begin{figure}[H]
	\centering
	\begin{subfigure}{0.65\textwidth}
		\centering
		\includegraphics[width=\textwidth, trim={0cm, 0cm, 0cm, 0cm},clip=true]{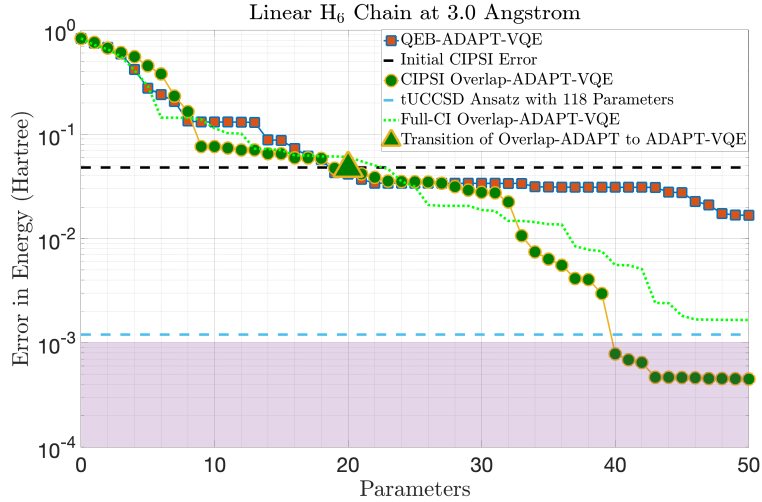} 
	\end{subfigure}
	\caption{Comparison of the CIPSI-Overlap-ADAPT-VQE and ADAPT-VQE for the ground state energy of a linear H$_6$ chain with an interatomic distance of 3 Angstrom. The plot represents the energy convergence as a function of the number of parameters in the ansatz. The CIPSI-Overlap ansatz is grown up to 20 parameters and then used as the initial state for an ADAPT-VQE process. This transition from Overlap-ADAPT-VQE to classical ADAPT-VQE is denoted by the top-pointing triangle. The horizontal black dotted line corresponds to the energy error of the initial CIPSI target wave-function. The light blue dotted line corresponds to the energy of the tUCCSD method \cite{yordanov2021qubit}, which consists of an ansatz wave-function composed of 118 generalised excitation evolutions acting on a reference Hartree-Fock state. The green dotted line corresponds to an FCI-Overlap-ADAPT-VQE procedure which is plotted simply as a reference. The pink area indicates chemical accuracy at $10^{-3}$ Hartree.}
	\label{fig:H6}
\end{figure}

Figure \ref{fig:H6} shows the energy convergence plot of the two different ADAPT-VQE protocols on the stretched linear H$_6$ system. We observe a significant difference in the results, with chemical accuracy being achieved using only 40 parameters when the QEB-ADAPT-VQE procedure is initialized with the overlap-guided-CIPSI intermediate wave-function whereas while the classical ADAPT-VQE ansatz is nearly 15 times less accurate despite using 50 parameters. Additional calculations revealed that with the classical QEB-ADAPT-VQE protocol requires more than 150 parameters to achieve chemical accuracy \cite{yordanov2021qubit}. This massive performance gap demonstrates that the CIPSI wave-function initialization guides the ansatz construction in a manner that avoids an initial massive energy plateau which impedes the progress of classical QEB-ADAPT-VQE. 

Let us emphasize here that the initial CIPSI wave-function was composed of only 50 determinants and had an error larger than $10^{-2}$ Hartree, which suggests that even a low accuracy classically computed target wave-function for the overlap-guided algorithm is enough to improve the convergence of the subsequent QEB-ADAPT-VQE procedure. This observation is particularly important since it highlights the potential of applying this CIPSI-Overlap-ADAPT procedure to much larger systems with strong correlation where CIPSI approaches are not effective and are simply unable to achieve chemical accuracy. For such systems, we can envision computing a CIPSI wave-function at the limit of classical computational resources, using this non-chemically accurate CIPSI wave-function as a target for the overlap-guided adaptive algorithm, and initialising a subsequent QEB-ADAPT-VQE procedure on a quantum computer in order to obtain a final result with chemical accuracy.

\begin{figure}[H]
	\centering
	\begin{subfigure}{0.495\textwidth}
		\centering
		\includegraphics[width=\textwidth, trim={0cm, 0cm, 0cm, 0cm},clip=true]{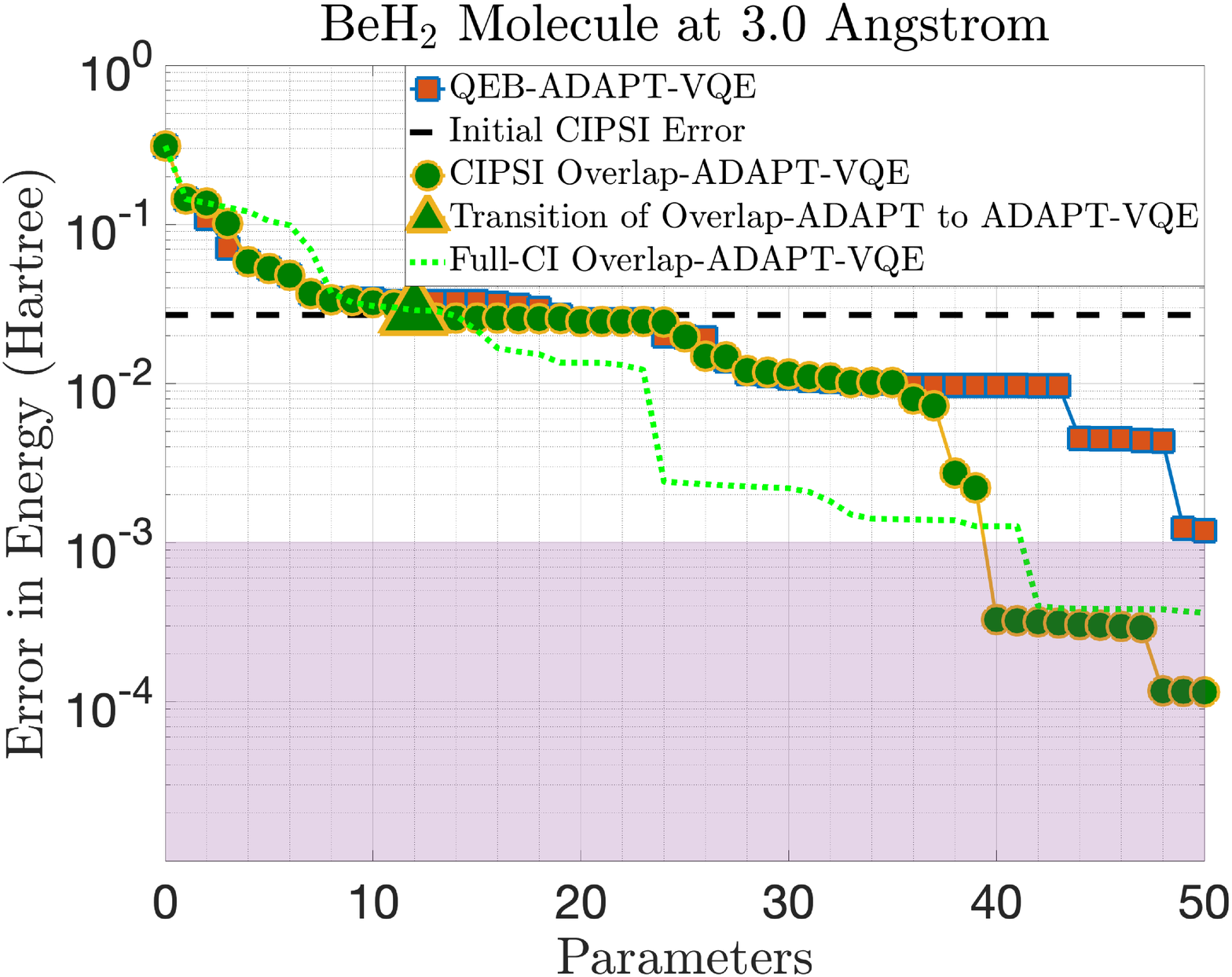} 
	\end{subfigure}\hfill
	\begin{subfigure}{0.495\textwidth}
		\centering
		\includegraphics[width=\textwidth, trim={0cm, 0cm, 0cm, 0cm},clip=true]{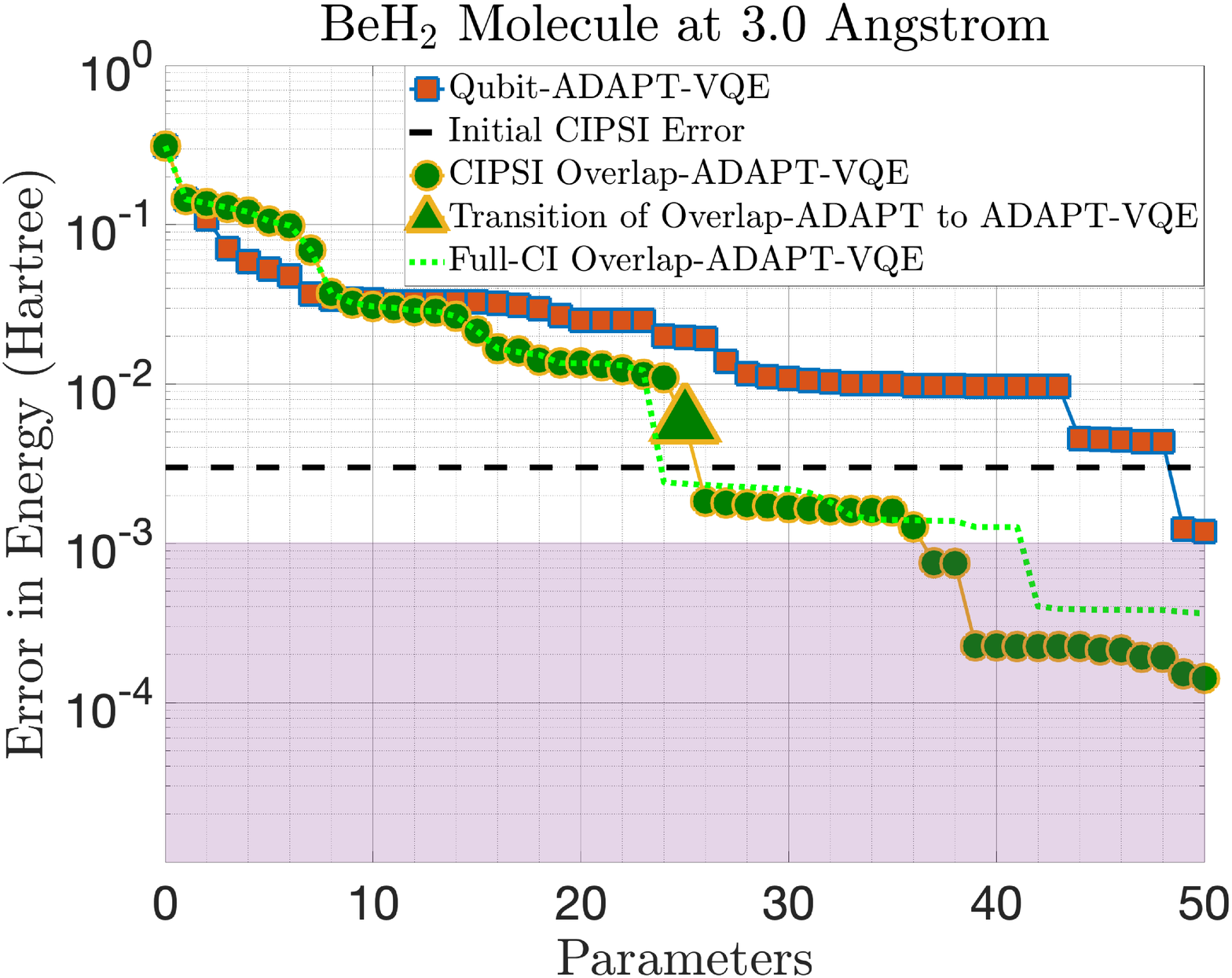} 
	\end{subfigure}
	\caption{Comparison of the CIPSI-Overlap-ADAPT-VQE and ADAPT-VQE for the ground state energy of a BeH$_2$ molecule with an interatomic distance of 3 Angstrom. The plots represent the energy convergence as a function of the number of parameters in the ansatz for two CIPSI initial wavefunction. The CIPSI-Overlap ansatz is grown up to 12 parameters (resp. 25 parameters) for the less accurate (resp. more accurate) initial CIPSI wave-function and then used as the initial state for an ADAPT-VQE process. This transition from Overlap-ADAPT-VQE to classical ADAPT-VQE is denoted by the top-pointing triangle. The horizontal dotted lines correspond to the energy error of the initial CIPSI target wave-function. The green dotted line corresponds to an FCI-Overlap-ADAPT-VQE procedure which is plotted simply as a reference. The pink area indicates chemical accuracy at $10^{-3}$ Hartree. }
	\label{fig:OG-BeH2}
\end{figure}

To further test the effectiveness of this CIPSI-Overlap-ADAPT approach, we return to the stretched BeH$_2$ molecule considered in the previous subsection. We employ two different CIPSI wave-functions as targets for the overlap-guided adaptive algorithm and use the approximate wave-functions obtained as high accuracy initializations for QEB-ADAPT-VQE procedures. Our results are displayed in Figure \ref{fig:OG-BeH2} and demonstrate that the CIPSI-Overlap-ADAPT produces a significantly more compact ansatz than the classical QEB-ADAPT-VQE procedure for both choices of CIPSI wave-functions. In both cases, the final accuracy of the wave-function with a maximal operator count of 50 operators is nearly an order of magnitude more than that of QEB-ADAPT-VQE. Furthermore, as noted in the case of the H$_6$ molecule, the choice of a low accuracy CIPSI wave-function as the initial target for the Overlap-ADAPT-VQE does not meaningfully degrade the final accuracy. Let us also remark here that the CIPSI-Overlap-ADAPT-VQE wave-function obtained at the end of the iterative process can then further be used a target for \emph{an additional} Overlap-ADAPT-VQE procedure, thereby further increasing the accuracy of the ansatz wave-function. In the case of the stretched BeH$_2$ molecule, this results in further minor improvements to the final energy that is achievable using a maximal operator count of 50, as displayed in Figure \ref{fig:OG-BeH2}.



%% file: Discussion.tex
In this study, we have explored the possibility of creating ansatz wave-functions for the variational quantum eigensolver that are more compact than the popular ADAPT-VQE at the chemical accuracy level for some small molecular systems. Since the overparametrization phenomenon observed in the ADAPT algorithm can be attributed to the algorithm's natural propensity to encounter local energy minima, we have proposed a new overlap-guided adaptative algorithm called Overlap-ADAPT-VQE, wherein the ansatz wave-function is grown by maximizing its overlap with an intermediate target wave-function that already captures some electronic correlation. We then use this overlap-guided ansatz as a high accuracy initialization for a classical ADAPT-VQE procedure. 

As a first test of our proposed approach, we used an existing ADAPAT-VQE ansatz wave-function as a target for the overlap-guided adaptive algorithm. The resulting ansatz wave-function was shown to achieve chemical accuracy using signficantly less operators than the classical ADAPT-VQE ansatz. We have also shown that this compression process can be carried out more than once and leads to an even more compact ansatz. For strongly correlated systems, the overlap-guided ansatz is noticeably steered by the target wave-function away from the majority of local traps that are typically encountered in standard ADAPT-VQE when starting from the Hartree-Fock state. While it appears that the ADAPT ansatz is already quite compact for systems with poor electronic correlation, the Overlap-ADAPT approach remains able to offer slight improvements.

Motivated next by the inability of ADAPT-VQE to process highly correlated systems such as the stretched linear H$_6$ chain using a reasonably compact ansatz, we combined classical selected-CI approaches and quantum computing by taking a CIPSI wave-function as a target for our overlap-guided adaptive algorithm. The resulting CIPSI-Overlap-ADAPT-VQE procedure produced a massive improvement over standard ADAPT-VQE, allowing us to reach chemical accuracy using an ansatz with only 40 operators compared to more than 150 for the classical ADAPT-VQE method. 

Previous studies have already investigated the use of additional classical computation to enhance the UCCSD or ADAPT-VQE methods and have demonstrated promising improvements \cite{filip2022reducing, romero2018strategies, metcalf2020resource, ryabinkin2020iterative, fedorov2022unitary}. Our work builds upon this research and contributes to this line of study. It is worth noting that the overlap-guided ansatz can also be interpreted as a state preparation algorithm for Hamiltonian simulation \cite{nielsen2002universal, abrams1999quantum, yao2021adaptive}, as it generates a state with high overlap on the ground state (see Figure \ref{fig:FCI_overlap}).  

However, within our new framework, the hybrid selected-CI-Overlap algorithm has the potential to bring a quantum advantage over classical quantum chemistry methods by following this procedure: pushing the classical computation of a complex molecular system to its limits, then generating the corresponding ansatz in a quantum computer using the Overlap adaptative algorithm, and further improving this ansatz through ADAPT-VQE and potentially additional overlap-guided compression steps. We are also testing the possibility of a final perturbative state (PT2) calculation following the spirit of the modern classical selected-CI approaches.

Finally, let us emphasise that Overlap-ADAPT-VQE is, by design, able to integrate seamlessly with the recent improvements made to ADAPT-VQE \cite{sapova2022variational, anastasiou2022tetris}, sharing the same structure and adaptive property while still leveraging its own unique approach to operator selection, and many combinations with ADAPT variants can now be proposed and studied. Conversely, convergence in overlaps can be achieved more quickly by incorporating a wider range of operators, such as generalized excitations or symmetry breaking operators, into the pool of operators used. This would lead to immediate improvements in the performance of the Overlap-ADAPT-VQE algorithm. To explore further the capabilities of the various Overlap-ADAPT approaches and their potential practical advantage over classical methods, we are currently working towards larger scale simulations on extended implementations encompassing larger qubit counts on present NISQ machines and new generation advanced simulators.




%% file: Appendix.tex
\subsection{Operator identification through overlap gradient computations}

Let us assume that we are at the $m^{\rm th}$ iteration of the ADAPT-VQE algorithm, i.e., we have a current ansatz wave-function $\ket{\Psi^{m-1}}$ as well as a target wave-function $\ket{\Psi_{\rm ref}}$. We now wish to identify the parameterized exponential qubit excitation evolution operator $\widehat{A}_m(\theta_m)$ whose action on the current ansatz $\ket{\Psi^{m-1}}$ will produce a new wave-function with the largest overlap with respect to the target wave-function using the overlap gradient expression given through Equation \eqref{eq:gradient}, i.e.,
\begin{equation*}
    \frac{\partial}{{\partial \theta_m}} \braket{\Psi_{\rm ref}|~\widehat{A}_m(\theta_m)\Psi^{m-1}}\big\vert_{\theta_m=0}.
\end{equation*}

To do so, recall that any parameterized qubit excitation evolution operator $\widehat{A}_m(\theta_m)$ is of the form (see Equation \eqref{eq:QEB_OP})
\begin{align*}
    \widehat{A}_m(\theta_m) = \exp\left(-i\theta \mathcal{B}_m\right),
\end{align*}
where $\mathcal{B}_m$ is a sum of so-called Pauli strings involving $X$ and $Y$ single qubit Pauli gates.

Using the chain rule, we therefore deduce that 
\begin{equation*}
    \frac{\partial}{{\partial \theta_m}} \braket{\Psi_{\rm ref}|~\widehat{A}_m(\theta_m)\Psi^{m-1}} =  \frac{\partial}{{\partial \theta_m}}\braket{\Psi_{\rm ref}|~\text{\rm exp}\left(-i\theta \mathcal{B}_m\right)\Psi^{m-1}} = \braket{\Psi_{\rm ref}|~i\mathcal{B}_m \text{\rm exp}\left(-i\theta \mathcal{B}_m\right)\Psi^{m-1}}.
\end{equation*}

Consequently, we obtain that
\begin{align}\nonumber
    \frac{\partial}{{\partial \theta_m}} \braket{\Psi_{\rm ref}|~\widehat{A}_m(\theta_m)\Psi^{m-1}}\big\vert_{\theta_m=0} &= \braket{\Psi_{\rm ref}|~i\mathcal{B}_m \text{\rm exp}\left(-i\theta \mathcal{B}_m\right)\Psi^{m-1}} \big\vert_{\theta_m=0}\\[1em] 
    &= \braket{\Psi_{\rm ref}|~i\mathcal{B}_m \Psi^{m-1}}. \label{eq:appendix_1}
\end{align}

We have thus reduced the problem of computing the sought-after overlap gradients at the $m^{\rm th}$ Overlap-ADAPT iteration to one of measuring the overlap between the target wave-function $\ket{\Psi_{\rm ref}}$ and an intermediate wave-function $\ket{i\mathcal{B}_m \Psi^{m-1}}$ where $\mathcal{B}_m$ is a sum of so-called Pauli strings involving $X$ and $Y$ single qubit Pauli gates. Let us emphasise here that our overlap operator identification process is based on finding qubit excitation evolution operator that maximises the \emph{magnitude} of the above computed gradient (and not the value of the gradient itself).

For classically represented target wave-functions (such as CIPSI wave-functions), such overlaps can be readily computed on classical architecture by making use of the usual determinant-based expansion of wave-functions. On the other hand, Equation \eqref{eq:appendix_1} cannot be directly evaluated on a quantum computer.

In order to evaluate Equation \eqref{eq:appendix_1} on quantum architecture, we proceed as follows. Using a direct calculation based on the Taylor series together with the commutation relations of Pauli matrices, we first deduce that
\begin{align*}
    \exp\left(-i\theta \mathcal{B}_m\right) &= I + (\cos(\theta)-1)\mathcal{B}_m^2 - i\sin(\theta) \mathcal{B}_m,
    \intertext{which immediately yields that}
    \braket{\Psi_{\rm ref}|~\widehat{A}_m(\theta_m)\Psi^{m-1}} &= \braket{\Psi_{\rm ref}|~\exp\left(-i\theta \mathcal{B}_m\right)\Psi^{m-1}}\\
    &= \braket{\Psi_{\rm ref}|\Psi^{m-1}}+ (\cos(\theta)-1)\braket{\Psi_{\rm ref}|\mathcal{B}_m^2\Psi^{m-1}} -\sin(\theta) \braket{\Psi_{\rm ref}|~i\mathcal{B}_m\Psi^{m-1}}.
\end{align*}

Plugging in the values $\theta_m = \pm\frac{\pi}{2}$ and $\theta_m = \pm \frac{\pi}{3}$, we can deduce from some tedious but straight-forward algebra that
\begin{align*}
   &\frac{1}{2} \left \vert\braket{\Psi_{\rm ref}|~\widehat{A}_m\left(-\frac{\pi}{2}\right)\Psi^{m-1}} \right\vert^2- \frac{1}{2} \left \vert\braket{\Psi_{\rm ref}|~\widehat{A}_m\left(\frac{\pi}{2}\right)\Psi^{m-1}}\right\vert^2\\[1em] 
   +&\frac{2}{\sqrt{3}} \left \vert\braket{\Psi_{\rm ref}|~\widehat{A}_m\left(\frac{\pi}{3}\right)\Psi^{m-1}}\right\vert^2 -\frac{2}{\sqrt{3}} \left \vert\braket{\Psi_{\rm ref}|~\widehat{A}_m\left(-\frac{\pi}{3}\right)\Psi^{m-1}}\right\vert^2\\[1em]
   =& -2 \braket{\Psi_{\rm ref}|\Psi^{m-1}} \braket{\Psi_{\rm ref}|~i\mathcal{B}_m\Psi^{m-1}},
\end{align*}
where we have used the fact that $\braket{\Psi_{\rm ref}|~i\mathcal{B}_m\Psi^{m-1}} \in \mathbb{R}$. As a consequence, we obtain that
\begin{align*}
    \left \vert \frac{\partial}{{\partial \theta_m}} \braket{\Psi_{\rm ref}|~\widehat{A}_m(\theta_m)\Psi^{m-1}}\big\vert_{\theta_m=0} \right \vert=\left \vert \braket{\Psi_{\rm ref}|~i\mathcal{B}_m\Psi^{m-1}} \right \vert &= \frac{1}{2}\frac{1}{\vert  \braket{\Psi_{\rm ref}|\Psi^{m-1}}\vert } \Bigg\vert \frac{1}{2} \left\vert\braket{\Psi_{\rm ref}|~\widehat{A}_m\left(-\frac{\pi}{2}\right)\Psi^{m-1}}\right \vert^2 \\
    &- \frac{1}{2} \left\vert\braket{\Psi_{\rm ref}|~\widehat{A}_m\left(-\frac{\pi}{2}\right)\Psi^{m-1}}\right \vert^2 +\frac{2}{\sqrt{3}} \left\vert\braket{\Psi_{\rm ref}|~\widehat{A}_m\left(\frac{\pi}{3}\right)\Psi^{m-1}}\right \vert^2\\ 
    &-\frac{2}{\sqrt{3}} \left\vert\braket{\Psi_{\rm ref}|~\widehat{A}_m\left(-\frac{\pi}{3}\right)\Psi^{m-1}}\right \vert^2\Bigg\vert.
\end{align*}

Each of the terms on the right-hand side of the above expression is now measurable on a quantum computer using the usual methods. Consequently, the sought-after overlap gradient can be computed at the cost of four measurements of the overlap $\braket{\Psi_{\rm ref}|~\widehat{A}_m(\theta_m)\Psi^{m-1}} $, i.e., one for each $\theta_m =\pm \frac{\pi}{2}$ and $\theta_m = \pm\frac{\pi}{3}$ as well as knowledge of the overlap from the previous iteration, i.e., $\vert  \braket{\Psi_{\rm ref}|\Psi^{m-1}}\vert $.\\

Let us remark here that since only four overlap evaluations are required to evalue a single overlap gradient, the Overlap-ADAPT-VQE process is very sober in terms of quantum resources. Indeed, we recall that computing a single expectation value of the Hamiltonian (or one energy gradient) involves measuring nearly each Pauli string in the Jordan-Wigner encoding of the Hamiltonian, which represents $O(N^4)$ function evaluations. Note however that some innovative techniques allow subsequent savings in the number of measurements for the expectation value of an Hamiltonian \cite{yen2023deterministic, loaiza2022reducing, choi2022improving}.

\vspace{5mm}

\newpage

\subsection{Additional data on operator and gate counts}
\begin{table}[h]
\centering
\caption{Single- and double-qubit evolution operator (resp. SQ, DQ) count for every simulation carried out in this paper. }
\label{tab:donnees}
\begin{tabular}{ccccc}
\hline
Molecule & Procedure & SQ operators & DQ operators & CNOTs \\
\hline
N$_2$ & All & 0 & 50 & 650 \\[1em]
BeH$_2$ at 1.3264\AA&  QEB-ADAPT & 8 & 42 & 570 \\
  & FCI-Overlap-ADAPT & 8 & 42 & 570 \\
  & QEB-ADAPT + Overlap-ADAPT & 8 & 42 & 570 \\[1em]
BeH$_2$ at 3\AA  & QEB-ADAPT & 7 & 43 & 580 \\
  & FCI-Overlap-ADAPT & 8 & 42 & 570 \\
  & 12 parameters CIPSI-Overlap-ADAPT & 6 & 44 & 590 \\
  & 25 parameters CIPSI-Overlap-ADAPT & 8 & 42 & 570 \\
  & 40 parameters QEB-ADAPT + Overlap-ADAPT & 6 & 34 & 460 \\
  & 45 parameters QEB-ADAPT + Overlap-ADAPT & 6 & 39 & 525 \\
  & 50 parameters QEB-ADAPT + Overlap-ADAPT & 7 & 43 & 580 \\[1em]
H$_6$ & QEB-ADAPT & 17 & 33 & 480 \\
   & FCI-Overlap-ADAPT & 0 & 50 & 650 \\
  & CIPSI-Overlap-ADAPT & 2 & 48 & 630 \\
\hline
\end{tabular}
\end{table}

%% file: main.bbl
\begin{thebibliography}{10}
\urlstyle{rm}
\expandafter\ifx\csname url\endcsname\relax
  \def\url#1{\texttt{#1}}\fi
\expandafter\ifx\csname urlprefix\endcsname\relax\def\urlprefix{URL }\fi
\expandafter\ifx\csname doiprefix\endcsname\relax\def\doiprefix{DOI: }\fi
\providecommand{\bibinfo}[2]{#2}
\providecommand{\eprint}[2][]{\url{#2}}

\bibitem{aspuru2005simulated}
\bibinfo{author}{Aspuru-Guzik, A.}, \bibinfo{author}{Dutoi, A.~D.},
  \bibinfo{author}{Love, P.~J.} \& \bibinfo{author}{Head-Gordon, M.}
\newblock \bibinfo{journal}{\bibinfo{title}{Simulated quantum computation of
  molecular energies}}.
\newblock {\emph{\JournalTitle{{S}cience}}} \textbf{\bibinfo{volume}{309}},
  \bibinfo{pages}{1704--1707} (\bibinfo{year}{2005}).

\bibitem{cao2019quantum}
\bibinfo{author}{Cao, Y.} \emph{et~al.}
\newblock \bibinfo{journal}{\bibinfo{title}{Quantum chemistry in the age of
  quantum computing}}.
\newblock {\emph{\JournalTitle{Chemical {R}eviews}}}
  \textbf{\bibinfo{volume}{119}}, \bibinfo{pages}{10856--10915}
  (\bibinfo{year}{2019}).

\bibitem{mcardle2020quantum}
\bibinfo{author}{McArdle, S.}, \bibinfo{author}{Endo, S.},
  \bibinfo{author}{Aspuru-Guzik, A.}, \bibinfo{author}{Benjamin, S.~C.} \&
  \bibinfo{author}{Yuan, X.}
\newblock \bibinfo{journal}{\bibinfo{title}{Quantum computational chemistry}}.
\newblock {\emph{\JournalTitle{Reviews of Modern Physics}}}
  \textbf{\bibinfo{volume}{92}}, \bibinfo{pages}{015003}
  (\bibinfo{year}{2020}).

\bibitem{peruzzo2014variational}
\bibinfo{author}{Peruzzo, A.} \emph{et~al.}
\newblock \bibinfo{journal}{\bibinfo{title}{A variational eigenvalue solver on
  a photonic quantum processor}}.
\newblock {\emph{\JournalTitle{{N}ature {C}ommunications}}}
  \textbf{\bibinfo{volume}{5}}, \bibinfo{pages}{1--7} (\bibinfo{year}{2014}).

\bibitem{mcclean2016theory}
\bibinfo{author}{McClean, J.~R.}, \bibinfo{author}{Romero, J.},
  \bibinfo{author}{Babbush, R.} \& \bibinfo{author}{Aspuru-Guzik, A.}
\newblock \bibinfo{journal}{\bibinfo{title}{The theory of variational hybrid
  quantum-classical algorithms}}.
\newblock {\emph{\JournalTitle{New {J}ournal of {P}hysics}}}
  \textbf{\bibinfo{volume}{18}}, \bibinfo{pages}{023023}
  (\bibinfo{year}{2016}).

\bibitem{bartlett1989coupled}
\bibinfo{author}{Bartlett, R.~J.}
\newblock \bibinfo{journal}{\bibinfo{title}{Coupled-cluster approach to
  molecular structure and spectra: a step toward predictive quantum
  chemistry}}.
\newblock {\emph{\JournalTitle{The {J}ournal of {P}hysical {C}hemistry}}}
  \textbf{\bibinfo{volume}{93}}, \bibinfo{pages}{1697--1708}
  (\bibinfo{year}{1989}).

\bibitem{romero2018strategies}
\bibinfo{author}{Romero, J.} \emph{et~al.}
\newblock \bibinfo{journal}{\bibinfo{title}{Strategies for quantum computing
  molecular energies using the unitary coupled cluster ansatz}}.
\newblock {\emph{\JournalTitle{Quantum Science and Technology}}}
  \textbf{\bibinfo{volume}{4}}, \bibinfo{pages}{014008} (\bibinfo{year}{2018}).

\bibitem{grimsley2018adapt}
\bibinfo{author}{Grimsley, H.~R.}, \bibinfo{author}{Economou, S.~E.},
  \bibinfo{author}{Barnes, E.} \& \bibinfo{author}{Mayhall, N.~J.}
\newblock \bibinfo{journal}{\bibinfo{title}{Adapt-vqe: An exact variational
  algorithm for fermionic simulations on a quantum computer}}.
\newblock {\emph{\JournalTitle{arXiv preprint arXiv:1812.11173}}}
  (\bibinfo{year}{2018}).

\bibitem{yordanov2021qubit}
\bibinfo{author}{Yordanov, Y.~S.}, \bibinfo{author}{Armaos, V.},
  \bibinfo{author}{Barnes, C.~H.} \& \bibinfo{author}{Arvidsson-Shukur, D.~R.}
\newblock \bibinfo{journal}{\bibinfo{title}{Qubit-excitation-based adaptive
  variational quantum eigensolver}}.
\newblock {\emph{\JournalTitle{Communications Physics}}}
  \textbf{\bibinfo{volume}{4}}, \bibinfo{pages}{1--11} (\bibinfo{year}{2021}).

\bibitem{tang2021qubit}
\bibinfo{author}{Tang, H.~L.} \emph{et~al.}
\newblock \bibinfo{journal}{\bibinfo{title}{{Qubit-Adapt-Vqe}: {A}n adaptive
  algorithm for constructing hardware-efficient ans{\"a}tze on a quantum
  processor}}.
\newblock {\emph{\JournalTitle{{PRX} {Q}uantum}}} \textbf{\bibinfo{volume}{2}},
  \bibinfo{pages}{020310} (\bibinfo{year}{2021}).

\bibitem{lee2018generalized}
\bibinfo{author}{Lee, J.}, \bibinfo{author}{Huggins, W.~J.},
  \bibinfo{author}{Head-Gordon, M.} \& \bibinfo{author}{Whaley, K.~B.}
\newblock \bibinfo{journal}{\bibinfo{title}{Generalized unitary coupled cluster
  wave functions for quantum computation}}.
\newblock {\emph{\JournalTitle{Journal of chemical theory and computation}}}
  \textbf{\bibinfo{volume}{15}}, \bibinfo{pages}{311--324}
  (\bibinfo{year}{2018}).

\bibitem{tilly2022variational}
\bibinfo{author}{Tilly, J.} \emph{et~al.}
\newblock \bibinfo{journal}{\bibinfo{title}{The variational quantum
  eigensolver: a review of methods and best practices}}.
\newblock {\emph{\JournalTitle{Physics Reports}}}
  \textbf{\bibinfo{volume}{986}}, \bibinfo{pages}{1--128}
  (\bibinfo{year}{2022}).

\bibitem{xie2022qubit}
\bibinfo{author}{Xie, Q.-X.}, \bibinfo{author}{Zhang, W.-g.},
  \bibinfo{author}{Xu, X.-S.}, \bibinfo{author}{Liu, S.} \&
  \bibinfo{author}{Zhao, Y.}
\newblock \bibinfo{journal}{\bibinfo{title}{Qubit unitary coupled cluster with
  generalized single and paired double excitations ansatz for variational
  quantum eigensolver}}.
\newblock {\emph{\JournalTitle{International Journal of Quantum Chemistry}}}
  \textbf{\bibinfo{volume}{122}} (\bibinfo{year}{2022}).

\bibitem{grimsley2022adapt}
\bibinfo{author}{Grimsley, H.~R.}, \bibinfo{author}{Barron, G.~S.},
  \bibinfo{author}{Barnes, E.}, \bibinfo{author}{Economou, S.~E.} \&
  \bibinfo{author}{Mayhall, N.~J.}
\newblock \bibinfo{journal}{\bibinfo{title}{{ADAPT-VQE} is insensitive to rough
  parameter landscapes and barren plateaus}}.
\newblock {\emph{\JournalTitle{arXiv preprint arXiv:2204.07179}}}
  (\bibinfo{year}{2022}).

\bibitem{zhao2022orbitaloptimized}
\bibinfo{author}{Zhao, L.} \emph{et~al.}
\newblock \bibinfo{title}{Orbital-optimized pair-correlated electron
  simulations on trapped-ion quantum computers} (\bibinfo{year}{2022}).
\newblock \eprint{2212.02482}.

\bibitem{gomes2021adaptive}
\bibinfo{author}{Gomes, N.} \emph{et~al.}
\newblock \bibinfo{journal}{\bibinfo{title}{Adaptive variational quantum
  imaginary time evolution approach for ground state preparation}}.
\newblock {\emph{\JournalTitle{Advanced Quantum Technologies}}}
  \textbf{\bibinfo{volume}{4}}, \bibinfo{pages}{2100114}
  (\bibinfo{year}{2021}).

\bibitem{jordan1993paulische}
\bibinfo{author}{Jordan, P.} \& \bibinfo{author}{Wigner, E.~P.}
\newblock \bibinfo{title}{{\"U}ber das {P}aulische {\"a}quivalenzverbot}.
\newblock In \emph{\bibinfo{booktitle}{The {C}ollected {W}orks of {E}ugene
  {P}aul {W}igner}}, \bibinfo{pages}{109--129} (\bibinfo{publisher}{Springer},
  \bibinfo{year}{1993}).

\bibitem{batista2001generalized}
\bibinfo{author}{Batista, C.~D.} \& \bibinfo{author}{Ortiz, G.}
\newblock \bibinfo{journal}{\bibinfo{title}{Generalized {J}ordan-{W}igner
  transformations}}.
\newblock {\emph{\JournalTitle{{P}hysical {R}eview {L}etters}}}
  \textbf{\bibinfo{volume}{86}}, \bibinfo{pages}{1082} (\bibinfo{year}{2001}).

\bibitem{nielsen00}
\bibinfo{author}{Nielsen, M.~A.} \& \bibinfo{author}{Chuang, I.~L.}
\newblock \emph{\bibinfo{title}{Quantum {C}omputation and {Q}uantum
  {I}nformation}} (\bibinfo{publisher}{Cambridge University Press},
  \bibinfo{year}{2000}).

\bibitem{kandala2017hardware}
\bibinfo{author}{Kandala, A.} \emph{et~al.}
\newblock \bibinfo{journal}{\bibinfo{title}{Hardware-efficient variational
  quantum eigensolver for small molecules and quantum magnets}}.
\newblock {\emph{\JournalTitle{Nature}}} \textbf{\bibinfo{volume}{549}},
  \bibinfo{pages}{242--246} (\bibinfo{year}{2017}).

\bibitem{o2016scalable}
\bibinfo{author}{O’Malley, P.~J.} \emph{et~al.}
\newblock \bibinfo{journal}{\bibinfo{title}{Scalable quantum simulation of
  molecular energies}}.
\newblock {\emph{\JournalTitle{Physical Review X}}}
  \textbf{\bibinfo{volume}{6}}, \bibinfo{pages}{031007} (\bibinfo{year}{2016}).

\bibitem{liu2019variational}
\bibinfo{author}{Liu, J.-G.}, \bibinfo{author}{Zhang, Y.-H.},
  \bibinfo{author}{Wan, Y.} \& \bibinfo{author}{Wang, L.}
\newblock \bibinfo{journal}{\bibinfo{title}{Variational quantum eigensolver
  with fewer qubits}}.
\newblock {\emph{\JournalTitle{Physical Review Research}}}
  \textbf{\bibinfo{volume}{1}}, \bibinfo{pages}{023025} (\bibinfo{year}{2019}).

\bibitem{sennane2023calculating}
\bibinfo{author}{Sennane, W.}, \bibinfo{author}{Piquemal, J.-P.} \&
  \bibinfo{author}{Ran{\v{c}}i{\'c}, M.~J.}
\newblock \bibinfo{journal}{\bibinfo{title}{Calculating the ground-state energy
  of benzene under spatial deformations with noisy quantum computing}}.
\newblock {\emph{\JournalTitle{Physical Review A}}}
  \textbf{\bibinfo{volume}{107}}, \bibinfo{pages}{012416}
  (\bibinfo{year}{2023}).

\bibitem{UCCSDT}
\bibinfo{author}{Haidar, M.}, \bibinfo{author}{Rančić, M.~J.},
  \bibinfo{author}{Maday, Y.} \& \bibinfo{author}{Piquemal, J.-P.}
\newblock \bibinfo{journal}{\bibinfo{title}{Extension of the {T}rotterized
  unitary coupled cluster to triple excitations}}.
\newblock {\emph{\JournalTitle{arXiv preprint arXiv:2212.12462}}}
  (\bibinfo{year}{2022}).

\bibitem{evangelista2019exact}
\bibinfo{author}{Evangelista, F.~A.}, \bibinfo{author}{Chan, G. K.-L.} \&
  \bibinfo{author}{Scuseria, G.~E.}
\newblock \bibinfo{journal}{\bibinfo{title}{Exact parameterization of fermionic
  wave functions via unitary coupled cluster theory}}.
\newblock {\emph{\JournalTitle{The {J}ournal of {C}hemical {P}hysics}}}
  \textbf{\bibinfo{volume}{151}}, \bibinfo{pages}{244112}
  (\bibinfo{year}{2019}).

\bibitem{yordanov2020efficient}
\bibinfo{author}{Yordanov, Y.~S.}, \bibinfo{author}{Arvidsson-Shukur, D.~R.} \&
  \bibinfo{author}{Barnes, C.~H.}
\newblock \bibinfo{journal}{\bibinfo{title}{Efficient quantum circuits for
  quantum computational chemistry}}.
\newblock {\emph{\JournalTitle{Physical Review A}}}
  \textbf{\bibinfo{volume}{102}}, \bibinfo{pages}{062612}
  (\bibinfo{year}{2020}).

\bibitem{PySCF}
\bibinfo{author}{Sun, Q.} \emph{et~al.}
\newblock \bibinfo{journal}{\bibinfo{title}{Pyscf: the python-based simulations
  of chemistry framework}}.
\newblock {\emph{\JournalTitle{WIREs Computational Molecular Science}}}
  \textbf{\bibinfo{volume}{8}}, \bibinfo{pages}{e1340} (\bibinfo{year}{2018}).

\bibitem{mcclean2020openfermion}
\bibinfo{author}{McClean, J.~R.} \emph{et~al.}
\newblock \bibinfo{journal}{\bibinfo{title}{Open{F}ermion: the electronic
  structure package for quantum computers}}.
\newblock {\emph{\JournalTitle{Quantum {S}cience and {T}echnology}}}
  \textbf{\bibinfo{volume}{5}}, \bibinfo{pages}{034014} (\bibinfo{year}{2020}).

\bibitem{STO3G}
\bibinfo{author}{Hehre, W.~J.}, \bibinfo{author}{Stewart, R.~F.} \&
  \bibinfo{author}{Pople, J.~A.}
\newblock \bibinfo{journal}{\bibinfo{title}{Self‐consistent
  molecular‐orbital methods. {I.} {U}se of {G}aussian expansions of
  {S}later‐type atomic orbitals}}.
\newblock {\emph{\JournalTitle{The Journal of {C}hemical {P}hysics}}}
  \textbf{\bibinfo{volume}{51}}, \bibinfo{pages}{2657--2664}
  (\bibinfo{year}{1969}).

\bibitem{2020SciPy-NMeth}
\bibinfo{author}{Virtanen, P.} \emph{et~al.}
\newblock \bibinfo{journal}{\bibinfo{title}{{{SciPy} 1.0: {F}undamental
  Algorithms for Scientific Computing in {P}ython}}}.
\newblock {\emph{\JournalTitle{{N}ature {M}ethods}}}
  \textbf{\bibinfo{volume}{17}}, \bibinfo{pages}{261--272}
  (\bibinfo{year}{2020}).

\bibitem{zhao2022orbital}
\bibinfo{author}{Zhao, L.} \emph{et~al.}
\newblock \bibinfo{journal}{\bibinfo{title}{Orbital-optimized pair-correlated
  electron simulations on trapped-ion quantum computers}}.
\newblock {\emph{\JournalTitle{arXiv preprint arXiv:2212.02482}}}
  (\bibinfo{year}{2022}).

\bibitem{GinSceCaf-CJC-13}
\bibinfo{author}{Giner, E.}, \bibinfo{author}{Scemama, A.} \&
  \bibinfo{author}{Caffarel, M.}
\newblock \bibinfo{journal}{\bibinfo{title}{Using perturbatively selected
  configuration interaction in quantum {M}onte {C}arlo calculations}}.
\newblock {\emph{\JournalTitle{Canadian Journal of Chemistry}}}
  \textbf{\bibinfo{volume}{91}}, \bibinfo{pages}{879--885}
  (\bibinfo{year}{2013}).

\bibitem{GinSceCaf-JCP-15}
\bibinfo{author}{Giner, E.}, \bibinfo{author}{Scemama, A.} \&
  \bibinfo{author}{Caffarel, M.}
\newblock \bibinfo{journal}{\bibinfo{title}{Fixed-node diffusion {M}onte
  {C}arlo potential energy curve of the fluorine molecule {F}$_2$ using
  selected configuration interaction trial wavefunctions}}.
\newblock {\emph{\JournalTitle{J. Chem. Phys.}}}
  \textbf{\bibinfo{volume}{142}}, \bibinfo{pages}{044115}
  (\bibinfo{year}{2015}).

\bibitem{hbci}
\bibinfo{author}{Holmes, A.~A.}, \bibinfo{author}{Tubman, N.~M.} \&
  \bibinfo{author}{Umrigar, C.~J.}
\newblock \bibinfo{journal}{\bibinfo{title}{Heat-bath configuration
  interaction: {A}n efficient selected configuration interaction algorithm
  inspired by heat-bath sampling}}.
\newblock {\emph{\JournalTitle{J. Chem. Theory Comput.}}}
  \textbf{\bibinfo{volume}{12}}, \bibinfo{pages}{3674--3680}
  (\bibinfo{year}{2016}).

\bibitem{SchEva-JCTC-17}
\bibinfo{author}{Schriber, J.~B.} \& \bibinfo{author}{Evangelista, F.~A.}
\newblock \bibinfo{journal}{\bibinfo{title}{{Adaptive Configuration Interaction
  for Computing Challenging Electronic Excited States with Tunable Accuracy}}}.
\newblock {\emph{\JournalTitle{Journal of {C}hemical {T}heory and
  {C}omputation}}} \textbf{\bibinfo{volume}{13}}, \bibinfo{pages}{5354--5366}
  (\bibinfo{year}{2017}).

\bibitem{GinTewGarAla-JCTC-18}
\bibinfo{author}{Giner, E.}, \bibinfo{author}{Tew, D.~P.},
  \bibinfo{author}{Garniron, Y.} \& \bibinfo{author}{Alavi, A.}
\newblock \bibinfo{journal}{\bibinfo{title}{Interplay between electronic
  correlation and metal--ligand delocalization in the spectroscopy of
  transition metal compounds: {C}ase study on a series of planar {C}u2$^+$
  complexes}}.
\newblock {\emph{\JournalTitle{Journal of Chemical Theory and Computation}}}
  \textbf{\bibinfo{volume}{14}}, \bibinfo{pages}{6240--6252}
  (\bibinfo{year}{2018}).

\bibitem{LooSceBloGarCafJac-JCTC-18}
\bibinfo{author}{Loos, P.-F.} \emph{et~al.}
\newblock \bibinfo{journal}{\bibinfo{title}{A mountaineering strategy to
  excited states: {H}ighly accurate reference energies and benchmarks}}.
\newblock {\emph{\JournalTitle{Journal of {C}hemical {T}heory and
  {C}omputation}}} \textbf{\bibinfo{volume}{14}}, \bibinfo{pages}{4360--4379}
  (\bibinfo{year}{2018}).

\bibitem{LooBogSceCafJac-JCTC-19}
\bibinfo{author}{Loos, P.~F.}, \bibinfo{author}{Boggio-Pasqua, M.},
  \bibinfo{author}{Scemama, A.}, \bibinfo{author}{Caffarel, M.} \&
  \bibinfo{author}{Jacquemin, D.}
\newblock \bibinfo{journal}{\bibinfo{title}{Reference energies for double
  excitations}}.
\newblock {\emph{\JournalTitle{Journal of {C}hemical {T}heory {C}omputation}}}
  \textbf{\bibinfo{volume}{15}}, \bibinfo{pages}{in press}
  (\bibinfo{year}{2019}).

\bibitem{QP2}
\bibinfo{author}{Garniron, Y.} \emph{et~al.}
\newblock \bibinfo{journal}{\bibinfo{title}{Quantum {P}ackage 2.0: {A}
  open-source determinant-driven suite of programs}}.
\newblock {\emph{\JournalTitle{J. Chem. Theory Comput.}}}
  \textbf{\bibinfo{volume}{15}}, \bibinfo{pages}{3591} (\bibinfo{year}{2019}).

\bibitem{ZhaLieuHof-JCTC-21}
\bibinfo{author}{Zhang, N.}, \bibinfo{author}{Liu, W.} \&
  \bibinfo{author}{Hoffmann, M.~R.}
\newblock \bibinfo{journal}{\bibinfo{title}{{Further Development of {iCIPT2}
  for Strongly Correlated Electrons}}}.
\newblock {\emph{\JournalTitle{Journal of {C}hemical {T}heory and
  {C}omputation}}} \textbf{\bibinfo{volume}{17}}, \bibinfo{pages}{949--964}
  (\bibinfo{year}{2021}).

\bibitem{benzene_bench}
\bibinfo{author}{Eriksen, J.~J.} \emph{et~al.}
\newblock \bibinfo{journal}{\bibinfo{title}{{The Ground State Electronic Energy
  of {B}enzene}}}.
\newblock {\emph{\JournalTitle{Journal of Physical Chemistry Letters}}}
  \textbf{\bibinfo{volume}{11}}, \bibinfo{pages}{8922--8929}
  (\bibinfo{year}{2020}).

\bibitem{LooDamSce-JCP-20}
\bibinfo{author}{Loos, P.-F.}, \bibinfo{author}{Damour, Y.} \&
  \bibinfo{author}{Scemama, A.}
\newblock \bibinfo{journal}{\bibinfo{title}{{The performance of {CIPSI} on the
  ground state electronic energy of {B}enzene}}}.
\newblock {\emph{\JournalTitle{Journal of {C}hemical {P}hysics}}}
  \textbf{\bibinfo{volume}{153}}, \bibinfo{pages}{176101}
  (\bibinfo{year}{2020}).

\bibitem{XuUejTen-JPCL-20}
\bibinfo{author}{Xu, E.}, \bibinfo{author}{Uejima, M.} \&
  \bibinfo{author}{Ten-no, S.~L.}
\newblock \bibinfo{journal}{\bibinfo{title}{{Towards Near-Exact Solutions of
  Molecular Electronic Structure: {F}ull Coupled-Cluster Reduction with a
  Second-Order Perturbative Correction}}}.
\newblock {\emph{\JournalTitle{Journal of {P}hysical {C}hemistry {L}etters}}}
  \textbf{\bibinfo{volume}{11}}, \bibinfo{pages}{9775--9780}
  (\bibinfo{year}{2020}).

\bibitem{GurDeuShePie-JCP-21}
\bibinfo{author}{Gururangan, K.}, \bibinfo{author}{Deustua, J.~E.},
  \bibinfo{author}{Shen, J.} \& \bibinfo{author}{Piecuch, P.}
\newblock \bibinfo{journal}{\bibinfo{title}{{High-level coupled-cluster
  energetics by merging moment expansions with selected configuration
  interaction}}}.
\newblock {\emph{\JournalTitle{Journal of {C}hemical {P}hysics}}}
  \textbf{\bibinfo{volume}{155}}, \bibinfo{pages}{174114}
  (\bibinfo{year}{2021}).

\bibitem{HurMalRan-JCP-73}
\bibinfo{author}{Huron, B.}, \bibinfo{author}{Malrieu, J.} \&
  \bibinfo{author}{Rancurel, P.}
\newblock \bibinfo{journal}{\bibinfo{title}{Iterative perturbation calculations
  of ground and excited state energies from multiconfigurational zeroth-order
  wavefunctions}}.
\newblock {\emph{\JournalTitle{{J}ournal of {C}hemical {P}hysics}}}
  \textbf{\bibinfo{volume}{58}}, \bibinfo{pages}{5745} (\bibinfo{year}{1973}).

\bibitem{three_class_CIPSI}
\bibinfo{author}{Evangelisti, S.}, \bibinfo{author}{Daudey, J.-P.} \&
  \bibinfo{author}{Malrieu, J.-P.}
\newblock \bibinfo{journal}{\bibinfo{title}{Convergence of an improved {CIPSI}
  algorithm}}.
\newblock {\emph{\JournalTitle{Journal of Chemical Physics}}}
  \textbf{\bibinfo{volume}{75}}, \bibinfo{pages}{91 -- 102}
  (\bibinfo{year}{1983}).

\bibitem{epstein}
\bibinfo{author}{Epstein, P.~S.}
\newblock \bibinfo{journal}{\bibinfo{title}{The {Stark} effect from the point
  of view of {S}chroedinger's quantum theory}}.
\newblock {\emph{\JournalTitle{Phys. Rev.}}} \textbf{\bibinfo{volume}{28}},
  \bibinfo{pages}{695--710} (\bibinfo{year}{1926}).

\bibitem{nesbet}
\bibinfo{author}{Nesbet, R.~K.}
\newblock \bibinfo{journal}{\bibinfo{title}{Configuration interaction in
  orbital theories}}.
\newblock {\emph{\JournalTitle{Proc. R. Soc. A}}}
  \textbf{\bibinfo{volume}{230}}, \bibinfo{pages}{312--321}
  (\bibinfo{year}{1955}).

\bibitem{filip2022reducing}
\bibinfo{author}{Filip, M.-A.}, \bibinfo{author}{Fitzpatrick, N.},
  \bibinfo{author}{Ramo, D.~M.} \& \bibinfo{author}{Thom, A.~J.}
\newblock \bibinfo{journal}{\bibinfo{title}{Reducing unitary coupled cluster
  circuit depth by classical stochastic amplitude prescreening}}.
\newblock {\emph{\JournalTitle{Physical {R}eview {R}esearch}}}
  \textbf{\bibinfo{volume}{4}}, \bibinfo{pages}{023243} (\bibinfo{year}{2022}).

\bibitem{metcalf2020resource}
\bibinfo{author}{Metcalf, M.}, \bibinfo{author}{Bauman, N.~P.},
  \bibinfo{author}{Kowalski, K.} \& \bibinfo{author}{De~Jong, W.~A.}
\newblock \bibinfo{journal}{\bibinfo{title}{Resource-efficient chemistry on
  quantum computers with the variational quantum eigensolver and the double
  unitary coupled-cluster approach}}.
\newblock {\emph{\JournalTitle{Journal of {C}hemical {T}heory and
  {C}omputation}}} \textbf{\bibinfo{volume}{16}}, \bibinfo{pages}{6165--6175}
  (\bibinfo{year}{2020}).

\bibitem{ryabinkin2020iterative}
\bibinfo{author}{Ryabinkin, I.~G.}, \bibinfo{author}{Lang, R.~A.},
  \bibinfo{author}{Genin, S.~N.} \& \bibinfo{author}{Izmaylov, A.~F.}
\newblock \bibinfo{journal}{\bibinfo{title}{Iterative qubit coupled cluster
  approach with efficient screening of generators}}.
\newblock {\emph{\JournalTitle{Journal of {C}hemical {T}heory and
  {C}omputation}}} \textbf{\bibinfo{volume}{16}}, \bibinfo{pages}{1055--1063}
  (\bibinfo{year}{2020}).

\bibitem{fedorov2022unitary}
\bibinfo{author}{Fedorov, D.~A.}, \bibinfo{author}{Alexeev, Y.},
  \bibinfo{author}{Gray, S.~K.} \& \bibinfo{author}{Otten, M.}
\newblock \bibinfo{journal}{\bibinfo{title}{Unitary selective coupled-cluster
  method}}.
\newblock {\emph{\JournalTitle{Quantum}}} \textbf{\bibinfo{volume}{6}},
  \bibinfo{pages}{703} (\bibinfo{year}{2022}).

\bibitem{nielsen2002universal}
\bibinfo{author}{Nielsen, M.~A.}, \bibinfo{author}{Bremner, M.~J.},
  \bibinfo{author}{Dodd, J.~L.}, \bibinfo{author}{Childs, A.~M.} \&
  \bibinfo{author}{Dawson, C.~M.}
\newblock \bibinfo{journal}{\bibinfo{title}{Universal simulation of hamiltonian
  dynamics for quantum systems with finite-dimensional state spaces}}.
\newblock {\emph{\JournalTitle{Physical Review A}}}
  \textbf{\bibinfo{volume}{66}}, \bibinfo{pages}{022317}
  (\bibinfo{year}{2002}).

\bibitem{abrams1999quantum}
\bibinfo{author}{Abrams, D.~S.} \& \bibinfo{author}{Lloyd, S.}
\newblock \bibinfo{journal}{\bibinfo{title}{Quantum algorithm providing
  exponential speed increase for finding eigenvalues and eigenvectors}}.
\newblock {\emph{\JournalTitle{Physical Review Letters}}}
  \textbf{\bibinfo{volume}{83}}, \bibinfo{pages}{5162} (\bibinfo{year}{1999}).

\bibitem{yao2021adaptive}
\bibinfo{author}{Yao, Y.-X.} \emph{et~al.}
\newblock \bibinfo{journal}{\bibinfo{title}{Adaptive variational quantum
  dynamics simulations}}.
\newblock {\emph{\JournalTitle{PRX Quantum}}} \textbf{\bibinfo{volume}{2}},
  \bibinfo{pages}{030307} (\bibinfo{year}{2021}).

\bibitem{sapova2022variational}
\bibinfo{author}{Sapova, M.~D.} \& \bibinfo{author}{Fedorov, A.~K.}
\newblock \bibinfo{journal}{\bibinfo{title}{Variational quantum eigensolver
  techniques for simulating carbon monoxide oxidation}}.
\newblock {\emph{\JournalTitle{{C}ommunications {P}hysics}}}
  \textbf{\bibinfo{volume}{5}}, \bibinfo{pages}{1--13} (\bibinfo{year}{2022}).

\bibitem{anastasiou2022tetris}
\bibinfo{author}{Anastasiou, P.~G.}, \bibinfo{author}{Chen, Y.},
  \bibinfo{author}{Mayhall, N.~J.}, \bibinfo{author}{Barnes, E.} \&
  \bibinfo{author}{Economou, S.~E.}
\newblock \bibinfo{journal}{\bibinfo{title}{{TETRIS-ADAPT-VQE}: {A}n adaptive
  algorithm that yields shallower, denser circuit ans\"atze}}.
\newblock {\emph{\JournalTitle{arXiv preprint arXiv:2209.10562}}}
  (\bibinfo{year}{2022}).

\bibitem{yen2023deterministic}
\bibinfo{author}{Yen, T.-C.}, \bibinfo{author}{Ganeshram, A.} \&
  \bibinfo{author}{Izmaylov, A.~F.}
\newblock \bibinfo{journal}{\bibinfo{title}{Deterministic improvements of
  quantum measurements with grouping of compatible operators, non-local
  transformations, and covariance estimates}}.
\newblock {\emph{\JournalTitle{npj Quantum Information}}}
  \textbf{\bibinfo{volume}{9}}, \bibinfo{pages}{14} (\bibinfo{year}{2023}).

\bibitem{loaiza2022reducing}
\bibinfo{author}{Loaiza, I.}, \bibinfo{author}{Marefat~Khah, A.},
  \bibinfo{author}{Wiebe, N.} \& \bibinfo{author}{Izmaylov, A.~F.}
\newblock \bibinfo{journal}{\bibinfo{title}{Reducing molecular electronic
  hamiltonian simulation cost for linear combination of unitaries approaches}}.
\newblock {\emph{\JournalTitle{Quantum Science and Technology}}}
  (\bibinfo{year}{2022}).

\bibitem{choi2022improving}
\bibinfo{author}{Choi, S.}, \bibinfo{author}{Yen, T.-C.} \&
  \bibinfo{author}{Izmaylov, A.~F.}
\newblock \bibinfo{journal}{\bibinfo{title}{Improving quantum measurements by
  introducing “ghost” pauli products}}.
\newblock {\emph{\JournalTitle{Journal of Chemical Theory and Computation}}}
  \textbf{\bibinfo{volume}{18}}, \bibinfo{pages}{7394--7402}
  (\bibinfo{year}{2022}).

\end{thebibliography}
